\documentclass[useAMS,usegraphicx,usenatbib]{mn2e}

\def\eg{{\it e.g.}}
\def\ie{{\it i.e.}}
\def\degrees{^\circ}
\def\kmsM{km s$^{-1}$ Mpc$^{-1}$}
\def\aj{AJ}
\def\apj{ApJ}
\def\apjl{ApJ}
\def\apjs{ApJS}
\def\aap{A\&A}
\def\mnras{MNRAS}
\def\nat{Nature}

\title[Double barred galaxies at intermediate redshifts: A feasibility study]{Double barred galaxies at intermediate redshifts:\\ A feasibility study}
\author[T. Lisker, V. P. Debattista, I. Ferreras and P. Erwin]{
Thorsten Lisker$^{1}$\thanks{E-mail: tlisker@astro.unibas.ch},
Victor P. Debattista$^{2}$\thanks{Brooks Prize Fellow},
Ignacio Ferreras$^{3}$
and Peter Erwin$^{4}$
\\
$^{1}$Astronomical Institute, Dept.\ of Physics and Astronomy, University of Basel, 
Venusstrasse 7, CH-4102 Binningen, Switzerland\\
$^{2}$Astronomy Department, University of Washington, Box 351580,
Seattle WA, 98195, USA\\
$^{3}$Department of Physics and Astronomy,
University College London, Gower Street, London WC1E 6BT, United Kingdom\\
$^{4}$Max-Planck-Institut f\"ur Extraterrestrische
Physik, Giessenbachstrasse, D-85748 Garching, Germany
}

\begin{document}

\date{Accepted ---. Received ---; in original form ---}

\pagerange{\pageref{firstpage}--\pageref{lastpage}} \pubyear{----}
  
\maketitle

\label{firstpage}

\begin{abstract}
Despite the increasing number of studies of barred galaxies at
intermediate and high redshifts, double-barred (S2B) systems have only
been identified in the nearby ($z\le 0.04$) universe thus far. In this
feasibility study we demonstrate that the detection and analysis of
S2Bs is possible at intermediate redshifts ($0.1\la z \la 0.5$) with
the exquisite resolution of the Hubble Space Telescope Advanced Camera
for Surveys ({\it HST}/ACS). We identify barred galaxies in the {\it
HST}/ACS data of the Great Observatories Origins Deep Survey (GOODS)
using a novel method.  The radial profile of the Gini coefficient -- a
model-independent structure parameter -- is able to detect bars in
early-type galaxies that are large enough that they might host an
inner bar of sufficient angular size. Using this method and subsequent
examination with unsharp masks and ellipse fits we identified the two
most distant S2Bs currently known (at redshifts $z=0.103$ and
$z=0.148$). We investigate the underlying stellar populations of these
two galaxies through a detailed colour analysis, in order to
demonstrate the analysis that could be performed on a future sample of
intermediate-redshift S2Bs.  We also identify two S2Bs and five S2B
candidates in the {\it HST}/ACS data of the Cosmic Evolution Survey
(COSMOS). Our detections of distant S2Bs show that deep surveys like
GOODS and COSMOS have the potential to push the limit for S2B
detection and analysis out by a factor of ten in redshift and look-back
time ($z\approx0.5$, $\Delta t\approx 5\,\rm{Gyr}$) compared to the
previously known S2Bs.  This in turn would provide new insight into
the formation of these objects.
\end{abstract}


\section{Introduction and motivation}
\label{sec:intro}

The high redshift universe allows us to probe galaxy evolution
directly, by presenting us with examples of galaxies at a fraction of
the age of the universe.  The resolution needed to study these objects
is possible either with ground-based adaptive optics or with the {\it
Hubble Space Telescope} ({\it HST}).  Using such observations, disk
galaxies have been detected up to redshift $z \sim 1$, when the
universe was not even half of its present age.  How do the properties
of these galaxies compare with those of local disks?  \citet{lil98}
showed that the scale-lengths of the $z\sim 1$ disks were similar to
those of local galaxies. Likewise, bulges at $z \sim 0.5$ are found
to have had a small amount of luminosity evolution and no significant
size evolution \citep{gla02}, although \citet{sim99} found a
population of $z\sim 0.5$ galaxies which have higher bulge surface
brightness than the local galaxies.

Morphologies of high-redshift galaxies also have been studied and
compared with those of local galaxies.  The most prominent
morphological feature of a large fraction of local galaxies is a bar
\citep{sw93review,esk00,kna00}.  Since
these can drive very strong secular evolution
\citep[\eg][]{lyn72,deb05b}, knowing how common they are at high $z$
is important for constraining disk evolution.  Using a sample of 46
galaxies observed in the Hubble Deep Field, \citet{abr99} concluded
that the fraction of barred galaxies was $\sim 24\%$ at $z\sim
0.2-0.7$ but decreased sharply from there to $z\sim 1.1$.  This result
subsequently turned out to be an artifact of the still limited
resolution of the Wide Field Planetary Camera 2 (WFPC2), and several
groups have now found a nearly constant barred fraction out to $z\sim
1$ using the higher resolution made possible by ACS
\citep{elm04,jog04}.  Other disk structures which have been identified
at high redshift include truncations at $0.6 < z < 1.0$
\citep{per04,tru05} and warps \citep{res02}.

Double-barred (S2B) galaxies, in which a small 'secondary' bar is
nested inside a larger 'primary' bar, have instead been identified
only out to $z=0.04$ thus far \citep{erw04}.  Indeed, the S2B catalog
of \citet{erw04} contains only two galaxies (out of 50) at a distance
of more than $100\,\rm{Mpc}$ and none further than $150\,\rm{Mpc}$.
S2Bs, which were first described by \citet{dv75}, have been postulated
to be a mechanism for driving gas past the inner Lindblad resonance of
primary bars to feed the supermassive black holes (SMBHs) that power
active galactic nuclei \citep{shl90}. Despite the possibility of their
playing such an important role in SMBH nurture, understanding them has
progressed significantly only in recent years.
Secondary bars have been sought in {\it gas-poor} early-type barred
galaxies with high resolution {\it HST} imaging in order to
distinguish secondary bars from other nuclear structures such as
disks, rings, spirals and dust extinction \citep{erw99,erw02}. We now
know that secondary bars are unambiguously found in about $30\%$ of
early-type barred galaxies.  Moreover, we finally have direct
observational proof for a kinematic decoupling between the primary and
secondary bars in one S2B galaxy, NGC 2950 \citep{cor03}. Such
decoupling is necessary, but not sufficient, for secondary bars to
funnel gas inwards.  On the theoretical side, important progress has
come from the development by \citet{mac97,mac00} of the orbit
formalism necessary for understanding their dynamics.  Finally,
simulators have started to produce gas-free S2B systems in their
simulations \citep{rau02}.  The secondary bars in these simulations
rotated faster than the primary bars and survived for several
gigayears, properties which favor SMBH fueling.

Observing S2Bs at significantly earlier epochs would allow us to
compare them with local S2Bs and would thus provide new insight into
the formation and evolution of these objects.  Because of their small
scales, their dynamical timescales may be quite short ($\sim 10$
Myr). However, this does not necessarily imply a short
\emph{lifetime}: since as many as $\sim 1/3$ of early-type barred
galaxies harbor secondary bars \citep[e.g.][]{erw02,lai02}, it appears very
unlikely that nuclear bars are 
destroyed and regenerated repeatedly in these nearly gas-free
systems. Instead, they probably have been in place a long time. It is
thus worthwhile to investigate whether the relative properties of the
secondary and primary bars show any evidence for evolution.  A
comparison of the stellar populations of inner and outer bar would be
one way to address this issue. If the stellar populations of secondary
bars are systematically younger than those of the corresponding
primary bars, this would support the hypothesis that nuclear bars form
out of gas driven in by the primary bar \citep{shl89}. If instead the
secondary bars consisted of older stellar populations than the primary bars,
this would make this scenario implausible.

Given that the difference between the optical properties of two
stellar populations decreases rapidly with age \citep[e.g.][]{bc03},
it would be desirable to observe S2B galaxies at larger look-back
times.  Moreover, a large enough sample of S2Bs at higher redshifts
would reveal a possible evolution of the ratios of
secondary-to-primary sizes. If inner bars lose angular momentum to the
outer bar, they might grow larger in radius, leading to a larger ratio
as time goes on.

While such questions motivate the search for and study of S2Bs
beyond the nearby universe, it is not yet clear whether such a study
would be at all possible with present-day instrumentation. Therefore,
as a first step, it is necessary to perform a \emph{feasibility
study}, i.e.\ to show that a) S2Bs can be found at significantly
larger look-back times than known to date, and b) the stellar
populations of primary and secondary bars can be analyzed and
distinguished.  An estimate of the limiting redshift for detecting
inner bars with the {\it HST} Advanced Camera for Surveys (ACS) is
provided by \citet{she03}, who give a lower limit for a bar semi-major
axis of $2.5$ times the point-spread function (PSF) full-width at half-maximum
(FWHM). Adopting $1$ PSF FWHM $=0\farcs11$ and the
concordance cosmology ($H_0 = 70$ \kmsM{} and $\Omega_\Lambda = 1 -
\Omega_m = 0.7$, which we assume throughout), the resolution of the
{\it HST}/ACS would be sufficient to identify inner bars out to
$z\approx 0.25$ provided that their sizes are similar to those of
local nuclear bars from \citet{erw04}. This limit is illustrated in
the bottom panel of Fig.~\ref{fig:sizes} by the dotted vertical lines, and
is significantly larger than the redshifts of known S2B systems, $z\le
0.04$.

Motivated by the foregoing estimate, we present a feasibility study for
detecting and analyzing double-barred galaxies at intermediate
redshifts, including our identification of the two
most distant S2Bs known so far.
In Section \ref{sec:data} we describe briefly the publicly available
data used in this study. Section \ref{sec:morph} explains how the
objects were found, including a novel technique for identifying
bars. A structural analysis of primary and secondary bars is presented
in Section \ref{sec:bars}, followed by a colour and stellar population
analysis in Section \ref{sec:pop}. Several (candidate-)S2Bs in
the Cosmic Evolution Survey (COSMOS) are presented in Section
\ref{sec:cosmos}, and a final discussion is presented in Section
\ref{sec:discuss}.


\section{The data}
\label{sec:data}

The {\it HST}/ACS images of the Great Observatories Origins Deep
Survey (GOODS) release v1.0 \citep{goods} cover an area of 300
arcmin$^2$ divided into two fields: the Hubble Deep Field North (HDFN)
and the Chandra Deep Field South (CDFS). They consist of four deep
images through the F435W ($B$), F606W ($V$), F775W ($i$), and F850LP
($z$) passbands of ACS. The CDFS is divided into 18 sections and the
HDFN into 17 sections with the images drizzled with a pixel scale of
$30$ mas, chosen to optimize the sampling of the PSF. The images are
not perfectly background subtracted: there is a remaining background
of the order of one third of the noise RMS, which can vary by up to a
factor of two across the image. Since this can contaminate
measurements of colour and morphological parameters, we subtracted an
individual background flux value for each object and band.  We defined
this background as the median pixel value within a box of $21\arcsec
\times 21\arcsec$ centered on the galaxy, after masking all sources.


\section{Object selection: Gini profiles}
\label{sec:morph}

Despite the common use of ellipse fits for bar identification, their
limitations are well known \citep[e.g.][]{erw03} and mainly originate
from the fact that ellipse fits assume a definite (namely elliptical)
isophotal shape of the object in question, which might not be the case
in reality. Motivated by the successful application of
model-independent structural parameters such as asymmetry or
concentration to galaxy morphology \citep[e.g.][]{abr03,con03}, we
sought a bar identification method that would rely solely on such
parameters and would therefore provide a computationally quick and
easy way to detect bars.

Given the usual difficulties in reliably identifying nuclear
structures in late-type galaxies because of the significant amount of
patchy obscuration that could mimic bar-like central shapes, we
restricted our study to early-type galaxies,
i.e.\ E/S0 galaxies as well as early-type spirals.
We did this by selecting
objects with central concentration values $C\ge 3.5$ \citep[cf.\
Fig.~11 of][]{lot04}, where $C = 5 \log[r(80\%)/r(20\%)]$ and
$r(x\%)$ is the radius enclosing $x\%$ of the total flux
\citep{ber00}. The total flux was measured in the $i$-band within a circular aperture
of $r = 1.5\times$ the Petrosian radius, $\rm{R_P}$ \citep{pet76},
where $\rm{R_P}$ is given by
\begin{equation}
\frac{\mu(\rm{R_P})}{\langle\mu \rangle(r<\rm{R_P})} = 0.2
\end{equation}
and $\mu$ is the surface brightness. We set cutoffs at apparent
magnitude $m_{\rm i,\,AB}\le 22.0$ and half-light radius $\geq 10$
pixels to ensure that the objects were bright and large enough for
identification of an inner bar.  These criteria yielded a working
sample of 231 objects in the GOODS dataset.

To detect large bars with major axis exceeding $15 - 20\%$ of the size
of the galaxy, we computed model-independent radial profiles of
structural parameters in circular apertures ranging from
$1/30\,\rm{R_P}$ to $1.5\,\rm{R_P}$ on the $i$-band images.
We focus on large bars for two reasons: first, the bar needs to stand
out in the radial structure profile, hence requiring its size to be a
significant fraction of the galaxy radius, and second, small bars are
less promising candidates for hosting nuclear bars that are large
enough to be detected.

One structure parameter -- the Gini coefficient, $G$ \citep{gini,abr03}
-- proved to be a useful diagnostic for large bars. We calculate $G$
following the definition of \citet{lot04},
\begin{equation}
G = \frac{ \sum^{N_{pix}}_i (2i -
N_{pix} -1) |X_i|}{N_{pix} (N_{pix}-1) \sum^{N_{pix}}_i |X_i|}
\end{equation}
where $N_{pix}$ is the number of pixels in the image and $X$ is the
flux and $X_1 \leq X_2 \leq X_3 \leq ... \leq X_{N_{pix}}$.  As
discussed by \citet{lot04}, the absolute values of $X$ are required in
order to preserve the correct structure at low flux levels, where
noise can result in negative values of $X$ after background
subtraction.  $G$ quantifies the distribution of flux values among the
pixels of an object's image.  The two extreme values are $G=0$ if all
pixels have the same flux, and $G=1$ if all the flux is concentrated
in a single pixel.  $G$ can therefore be thought of as measuring the
contrast between the faintest and the brightest regions of a galaxy's
image, \emph{no matter where in the galaxy they are located}; thus it
does not assume any shape for the object.  In our definition $G(r)$ is
computed within each aperture as described above.  Circular apertures
are preferred over elliptical ones because the shape of a bar clearly
stands out in a circular aperture. An elliptical aperture instead
would fit the bar's shape to a higher degree, preventing the bar from
standing out quite as much.

What do the Gini profiles of ordinary galaxies look like? From the
objects in our working sample we assembled a subsample of such
galaxies, ranging from ellipticals to late-type spirals and irregulars
based on a visual classification. Since we selected objects with high
concentration values, late-type spirals appear in our working sample
generally when they are seen nearly edge-on. Similarly, the objects
classified 'irregular' still require a large amount of their light to
be centrally concentrated to enter the sample. It is apparent from
Fig.~\ref{fig:mopro} that $G(r)$ steadily increases with radius for
E/S0 galaxies (possibly including edge-on early-type spirals), while
for spirals it reaches a maximum and decreases thereafter. Late-type
spirals can be distinguished because their maxima in $G(r)$ are lower
and are reached earlier than in early-type spirals

In the Gini profile of several galaxies, a plateau is seen at {\it
intermediate} radii (top left panel of Fig.~\ref{fig:mopro}). We
identify such a feature by its mathematical properties: the second
derivative changes from negative to positive, with the first
derivative positive before and after, but becoming small, or possibly
negative, in between. We approximate the first and second derivative
by considering small ($\rm{R_P}/30$) intervals, which are then
smoothed by averaging over three data points at each step.
 An object was selected if its profile shows a plateau that is at least
 as strong as the one presented in the upper right panel of
 Fig.~\ref{fig:mopro}. This also requires the bar to be 'strong', i.e.\
 a significant fraction of the total light needs to make up the bar
 shape. Bars that are so weak that they would hardly be seen without
 the help of unsharp masking or ellipse fits probably would not have enough
 effect on the Gini coefficient to produce a clear plateau in the
 profile.
A
plateau is evident in 46 objects of our working sample.
Visual
inspection shows that 17 of them indeed have a large bar, 23 galaxies
are unbarred, while for 6 galaxies we could not decide whether a bar
was present based on visual inspection.

How can the shape of $G(r)$ for a barred galaxy be understood? First,
the initial increase of $G(r)$ with radius is caused by the brightest
pixels being all located in the central region; thus increasing the
radius adds only fainter pixels.  This leads to a continuously
increasing spread in the pixel flux values, and thus each Gini profile
begins with increasing $G(r)$, independent of galaxy type.  However, a
bar is then able to maintain a relatively high surface brightness out
to a larger radius than an unbarred object, therefore the fraction of
faint pixels remains lower, which probably leads to the plateau in
$G(r)$.  Thus the Gini profile provides a simple way of identifying
large, strong bars, and it can be computed easily and automatically
for all sources detected in a deep survey such as GOODS.

Although we did successfully apply this bar detection method to the
GOODS dataset, we point out that it is not yet optimized: the fraction
of unbarred objects among the selected galaxies is $\ge 50\%$, which
would seem to not be a significant improvement over ellipse fits.
Moreover, when visually inspecting all 231 galaxies of our working
sample, 11 objects possess a bar but were missed by our method, partly
because their bar is relatively small, but also because of spirals and
lopsidedness. A quick visual study revealed that the method could be
improved further, for example by including a second parameter to
exclude those systems which are asymmetric. Moreover, our above
interpretation of the cause of the plateau in the profile might be too
simplistic: it could well be possible that not (only) the bar itself
is causing the plateau, but rather the configuration of a barred
galaxy as a whole, i.e.\ the distribution of pixel flux values
\emph{around} the bar. A full study of these issues is beyond the
scope of this paper and will be addressed elsewhere (Pasquali et al.,
in prep.). Since completeness is not the goal for our
feasibility study, we are satisfied with the easy derivation of the
Gini profiles and defer further quantitative evaluation of the
method's potential. 

Can the method be applied recursively to identify nuclear bars inside
barred galaxies?  This did not prove feasible due to the small
apparent size of nuclear bars and, more importantly, the significant
amount of smoothly distributed underlying galaxy light which decreases
the contrast of nuclear bars (for the same reason, unsharp masks are
such a useful tool for studying nuclear bars).
Thus we had to inspect the selected barred objects visually and
through unsharp masks in order to identify nuclear bars (Sect.\
\ref{sec:bars}). Two objects were clearly identified as double-barred:
GOODS J033230.93-273923.7, hereafter S2B\,1, and GOODS
J033233.46-274312.8, hereafter S2B\,2; both lie in the CDFS part of
the GOODS survey.


\section{Structural Properties}
\label{sec:bars}

Here we present a structural analysis of S2B\,1 and S2B\,2.  For this
purpose we use unsharp masks and ellipse fits as in \citet{erw04}.

\subsection{S2B\,1 -- a double-barred galaxy at $z=0.148$}
\label{sec:barry}

The left panels of Fig.~\ref{fig:images} show, from top to bottom, an
$i$-band image of S2B\,1, an unsharp mask, and the ellipse fits.  The
unsharp mask was created by dividing the original image with one
convolved with a Gaussian of $\sigma=40$ pixels; in the center we
insert another unsharp mask created through a convolution with a much
smaller Gaussian ($\sigma=4$ pixels) to highlight the inner structure.
Apart from the huge primary bar, a very prominent secondary bar is
present along with a nuclear spiral.  Its spectroscopic redshift, as
measured by the VIMOS VLT Deep Survey \citep{vvds}, is $z=0.148$.
S2B\,1 is therefore the most distant S2B galaxy known, at a luminosity
distance of
$703\,\rm{Mpc}$\footnote{Distances and image scales were derived with
  Ned Wright's Cosmology Calculator,\\
{\tt http://www.astro.ucla.edu/$\sim$wright/CosmoCalc.html}}.
The black scale bar in the image
corresponds to a physical length of $5\,\rm{kpc}$; the image scale is
$2.585$ kpc/$\arcsec$. We
visually classify S2B\,1 as SB0/a.

In order to quantify the structure of S2B\,1, we performed an ellipse
fit using the IRAF task {\sc ELLIPSE}.  The result is shown in
Fig.~\ref{fig:images}; both the primary and secondary bar can be
clearly 
identified as significant local peaks in the ellipticity.  We did not
attempt a deprojection of S2B\,1, since its outer structure is
difficult to determine: the primary bar dominates the galaxy's
appearance, and it cannot be ascertained whether it is circular or
still part of an oval distortion. Since the primary bar lies along the
major axis of the outer isophotes, as indicated by the nearly constant
position angle in the ellipse fits, we do not expect measurements of
its size and shape to be significantly affected by a, probably modest,
inclination.

Following \citet{erw04}, we measured the primary bar's semi-major axis
at its maximum ellipticity $a_{\epsilon,1}=7.0\,\rm{kpc}$, and
the upper limit on the bar semi-major axis as
$L_{bar,1}=8.3\,\rm{kpc}$, where $L_{bar}$ is the smaller of $a_{\rm
min}$ and $a_{10}$, with $a_{\rm min}$ being the semi-major axis at
the first ellipticity minimum outside the bar, and $a_{10}$ being the
semi-major axis where the position angle has changed by $10 \degrees$
from the bar position angle.  The equivalent values for the secondary
bar are $a_{\epsilon,2}=0.8\,\rm{kpc}$ and
$L_{bar,2}=1.1\,\rm{kpc}$.  The physical size of the primary bar in
S2B\,1 is larger than 47 of the 50 S2Bs presented by \citet{erw04},
while the secondary is larger than $76\%$ of Erwin's sample.

Visual inspection of the immediate environment of S2B\,1 reveals
a jet-like object close to it and pointing towards its center
(Fig.~\ref{fig:barryouter}). This object is seen in all bands. If the
secondary bar in S2B\,1 is feeding gas to an SMBH, then a jet
originating from the SMBH may indeed be expected.  However, its
photometric redshift provided by the COMBO-17 survey \citep{combo17}
is $z=0.681$, suggesting the object is a background galaxy. On the
other hand, one would not expect a photometric redshift (which relies
on template spectral energy distributions) to yield useful results if
the object actually was a jet. The X-ray flux of the galaxy would
allow us to test for an active black hole in the center.
The Extended CDFS
survey \citep{vir06} does not detect S2B\,1, but it
provides an upper limit on the X-ray flux. The 3 sigma upper limit on
the count rate is $6.35\times10^{-5}\,\rm{counts/s}$ (Virani, priv.\
comm.), which translates into an upper limit on the flux in the soft
bands ($0.5 - 2.0\,\rm{keV}$) of $2.7\times
10^{-16}\,\rm{erg\,cm^{-2}\,s^{-1}}$, using the Chandra X-ray Center
Proposal Planning Toolkit PIMMS v3.7\footnote{{\tt
    http://cxc.harvard.edu/toolkit/pimms.jsp}} with a power law
spectral model and a photon index of 1.8. This corresponds to an upper
limit on the soft X-ray luminosity of $L_{\rm X} = 1.6\times
10^{40}\,\rm{erg\,s^{-1}}$. By applying equations 1 and 2 of
\citet{hor03} we derive an upper limit on the X-ray-to-optical flux
ratio of S2B\,1 of $\log(f_{\rm X}/f_{\rm R})\le -2.9$. This value lies
two orders of magnitude below the typical values for AGNs \citep[cf.\
  Fig.\ 3 of][]{hor03}; however, a heavily obscured AGN cannot be
excluded.

%
%
%

\citet{afo06} presented Australia Telescope Compact Array radio
(1.4\,GHz) observations of the CDFS. They found a total of 64 radio
sources within the {\it HST}/ACS region of the GOODS/CDFS with flux
densities between $63\,{\rm \mu Jy}$ and $20\,{\rm mJy}$. However,
S2B\,1 was not detected in this survey.

\subsection{S2B\,2 -- a double-barred galaxy at $z=0.103$}
\label{sec:mary}

The right panels of Fig.~\ref{fig:images} show an $i$-band image,
unsharp mask, and the ellipse fits for S2B\,2, which we classify as
SB0/a.  The unsharp masks were produced with Gaussians of $\sigma=1.5$
pixels and $\sigma=30$ pixels. The small unsharp mask reveals an inner
bar and spiral in the center of S2B\,2. This is confirmed by a local
peak in ellipticity, although this peak is relatively weak due to the
small apparent size of the bar. The position angle nicely tracks the
inner bar and then changes by about $90 \degrees$ outside of it,
since the inner and outer bars in this galaxy are nearly perpendicular.
S2B\,2's spectroscopic
redshift, as measured by 2dFGRS \citep{2df}, is $z=0.103$, putting it
at a luminosity distance of $475\,\rm{Mpc}$.  The black scale bar in
the image corresponds to a physical length of $5\,\rm{kpc}$; the image
scale is $1.893$ kpc/$\arcsec$.

We deprojected the galaxy by assuming that the ellipticity minimum
outside of the primary bar is the apparent ellipticity of the galaxy
as a whole; we then obtained the deprojected size of the inner bar as
$a_{\epsilon,2}=0.4\,\rm{kpc}$ and $L_{bar,2} =
0.5\,\rm{kpc}$. The corresponding values for the primary bar are
$a_{\epsilon,1} = 4.2\,\rm{kpc}$ and $L_{bar,1} = 6.4\,\rm{kpc}$,
which did not require deprojection since the primary bar is almost
along the galaxy's apparent major axis.

The primary bar of S2B\,2 is slightly smaller than the median size of
the 50 S2Bs in \citet{erw04}, while the secondary bar is smaller than
that of $70\%$ of the same sample.  With
$a_{\epsilon,2}=0\farcs17=1.5$ PSF FWHM the apparent size of the
secondary bar is obviously just large enough for it to be detected by
{\it HST}/ACS ($1$ PSF FWHM $=0\farcs11$ in the $i$-band and similar in
the other bands). We therefore conclude that $z
\sim 0.1$ would be the maximum redshift for {\it HST}/ACS detection of
the inner bar for the smallest $\approx 30\%$ of the currently known
S2Bs.

S2B\,2 has a faint (pseudo) ring around the main galaxy, resembling
the R$^\prime_1$ class of \citet{but91}.  Rings are quite
common in barred galaxies, and simulations show that they can be
produced at bar resonances \citep[see, \eg\, the review by][]{but96}.

%
%
%

S2B\,2 is not detected in the radio observations of
\citet{afo06}, but is detected in the CDFS survey of the Chandra X-Ray
Observatory \citep{ale03,gia02}. Its X-ray flux is $1.8\times
10^{-16}\,\rm{erg\,cm^{-2}\,s^{-1}}$ in the soft bands ($0.5 -
2.0\,\rm{keV}$), which translates into a soft X-ray luminosity of $L_{\rm X}
= 4.9\times 10^{39}\,\rm{erg\,s^{-1}}$. The upper limit on the flux in
the hard bands ($2.0 - 8.0\,\rm{keV}$) is $9.3\times
10^{-16}\,\rm{erg\,cm^{-2}\,s^{-1}}$ \citep{ale03}.
With an 
 X-ray-to-optical flux
ratio of $\log(f_{\rm X}/f_{\rm R})= -3.4$ S2B\,2
belongs to the optically bright, X-ray faint (OBXF) class of sources
\citep{hor03}.
These objects are thought to be quiescent, normal
galaxies, with their high-energy emission originating from stellar
processes or from low-level accretion onto SMBHs, but not from
luminous active galactic nuclei
unless they are heavily obscured.
Since the optical and X-ray positions
only differ by $0\farcs4$ -- which equals the median positional
uncertainty of the faint X-ray sources \citep{ale03} -- it is likely
to originate from the central region. While we cannot gain more
insight into the processes generating the X-ray emission of S2B\,2,
more detections of intermediate-redshift S2Bs at X-ray wavelengths
would test the hypothesis that gas in S2Bs is fed to SMBHs.


\section{Stellar population analysis}
\label{sec:pop}

Comparisons of the available photometry from the ACS images with
population synthesis models allow us to estimate the age and
metallicity distribution
of the stellar component in the two galaxies.  A pixel by pixel
approach has the advantage of spatially resolving regions which may
have undergone different histories of star formation. However, the
uneven distribution of the signal-to-noise ratio (SNR) across the
galaxy images results in complicated maps of uncertainties which are
hard to interpret. In order to improve the spatially resolved
analysis, we decided to perform an adaptive binning following the
Optimal Voronoi Tessellation method of \citet{cap03}, which has
already been applied successfully to a study of early-type galaxies in
GOODS/CDFS \citep{egoodss}. A Voronoi tessellation divides a region
starting from a distribution of points called {\sl generators}.  The
area (or volume) is then binned in those regions closest to a given
generator. Motivated by the analysis of integral field spectroscopy,
\citet{cap03} presented an algorithm optimised for the binning of a 2D
map into bins which yield a small scatter of the SNR with respect to
some target value and which are kept reasonably round in order to
preserve, as much as possible, the spatial resolution.  When we apply
this method to colour maps, we ensure that the subsequent stellar
population analysis is as homogeneous as possible.

We generated colour maps by registering the images and convolving them
with the PSF corresponding to the other band used for a given colour
-- \eg\ for a $V-i$ colour map we convolve the $i$-band image with the
$V$-band PSF and vice versa.  We measured the PSFs from a set of stars
in the GOODS/CDFS field.  A
recent analysis performed by the GEMS group \citep{rix04} found that a
universal PSF obtained from a set of bright but unsaturated stars is
fully sufficient for all GOODS sections and for all positions inside
each section. For the Optimal Voronoi Tessellation on the registered
and convolved images, we chose a target SNR of $20$ in the ratio of
the $V$ to $i$ fluxes, which amounts to an uncertainty of $\pm
0.05$~mag for the $V-i$ colour in each bin.  The same bins were used
in all other colours ($B-V$ and $i-z$); the average errors are $0.1$
mag in $B-V$ and $0.05$ mag in $i-z$.  Note that S2B\,1 does not have
a $B$-band image because it falls close to the edge of the field.
Figures~\ref{fig:BC} and \ref{fig:MC} show the colour maps for S2B\,1
and S2B\,2, respectively. The distribution of colours is somewhat
patchier than expected from the above errors only (\ie\ the RMS of the
colour values exceeds the errors), particularly in S2B\,1,
which is partly due to dust and partly to population differences.

Table~\ref{tab:phot} lists some integrated properties of our objects.
The $V$-band absolute luminosity was computed from the observed
passband which gives the smallest K correction, \ie\ F606W for both
S2B\,1 and S2B\,2, with a correction $K_{\rm AB}\sim 0.04$~mag.  The
spectral energy distribution for these corrections was obtained from a
simple stellar population with solar metallicity, whose age best
matches the observed colours.

\subsection{S2B\,1}
\label{sec:barrycol}

A young stellar component in the inner spiral structure is readily
visible in the $V-i$ colour map in Fig.~\ref{fig:BC}, at around
$0\farcs75$ from the center. Ongoing star formation is also deduced
from a low-resolution spectrum of S2B\,1 from the VIMOS VLT Deep
Survey of the CDFS region \citep{vvds}. The spectrum spans the
5600-9000\AA\ region and features H$\alpha$ and [O\,III] but no
[N\,II] emission, a signature of star-forming HII regions. Although
these might coincide with the above region of young stars, no spatial
resolution is possible due to the slit size ($1\arcsec \times
10\arcsec$) and to ground-based
resolution.

The age-metallicity degeneracy prevents us from getting an accurate
estimate of the age and metallicity distribution. Nevertheless, we can
test for gross differences between the stellar populations in the
primary and secondary bars via a colour-colour diagram.  The top panel
of Fig.~\ref{fig:ccd} shows the $V-i$ versus $i-z$ colour distribution
of S2B\,1.  Each ``pixel'' corresponds to a Voronoi-tessellated bin
with the targeted SNR of 20 in the ratio of $V$ to $i$, which aims at
a 0.05~mag photometric error in $V-i$. The actual tessellation
produced an average error of 0.047~mag with a 0.033~mag RMS scatter.
Regions in the inner bar are represented by filled dots. We selected
those regions of the galaxy with an $i$-band surface brightness
between 17.5 and 18.5~mag/arcsec$^2$, so that we exclude the very
inner region which appears to be affected by dust. The filled
triangles give the colours in the outer bar, located in the region
with surface brightness between 21 and 22~mag/arcsec$^2$. We select
pixels in a $20\degrees$ wedge aligned with the position angle of the
outer bar.  We also distinguish pixels in two radial bins: the light
grey symbols correspond to distances $r<1\farcs8$ from the center
while the dark grey symbols are pixels at $r>1\farcs8$.

For comparison, we show the colours of simple stellar populations from
the latest models of \citet{bc03}. Each line represents an age
sequence between 2 and 10 Gyr, and three metallicities are considered,
as labelled in the figure. The dashed lines are dustless populations,
whereas the solid lines assume $E(B-V)=0.2$ and a simple dust
reddening law \citep{cha00}.  The inner bar shows the bluest colours
in $i-z$, which can be interpreted as indicating that its metallicity
is lower than of the outer bar, and in addition either the stellar
ages or the amount of dust is higher.  However a unique interpretation
is not possible with these data.  The outer bar regions possess a
larger colour spread than the inner bar, which presumably is caused
mainly by dust (given the patchy distribution of colours discussed
above), but may partly be due to radial variations in the stellar
populations.

\subsection{S2B\,2}
\label{sec:marycol}

This galaxy has a conspicuous ring of young stars at a projected
distance $\sim 3.8$~kpc ($\sim 2\arcsec$) from the center
(Fig.~\ref{fig:MC}). Interior to this ring, the
red colours are indicative of either old populations or dust
extinction. The bottom panel of Fig.~\ref{fig:ccd} shows the
colour-colour diagram of S2B\,2. The inner region -- shown as dots --
corresponds to an $i$-band surface brightness between 17 and
18~mag/arcsec$^2$.  The filled triangles are pixels in the outer bar -- selected
in a $20\degrees$ wedge centered at the position angle of the bar
with surface brightness between 20 and 21~mag/arcsec$^2$.
Two radial bins are chosen in the same way as with S2B\,1, and similar
population synthesis model predictions are shown. Note the
remarkable
difference between the outer bar and the red colours of the inner bar,
which is significantly affected by dust. The figure hints at younger
ages in the outer parts of the primary bar, which is clearly due to the
ring of young stars mentioned above.

\section{S2Bs in COSMOS}
\label{sec:cosmos}

In the course of our study, the reduced {\it HST}/ACS $i$-band imaging
data (v1.2) of the first square degree of the Cosmic Evolution Survey
(COSMOS\footnote{See {\tt http://www.astro.caltech.edu/cosmos}})
became publicly available. Since these data cover an area about twelve
times larger than GOODS, they can be expected to host several S2B
galaxies at intermediate redshifts. The COSMOS image depth is only
slightly lower than that of GOODS, therefore a search for S2Bs with
our Gini profile method and subsequent visual inspection appeared
promising.
We applied the same selection criteria as for our GOODS dataset, i.e.\
$C\ge 3.5$, $m_{\rm i,\,AB}\le 22.0$, and half-light radius $\geq 10$
pixels.
Indeed, we were able to identify two S2Bs and five S2B
candidates which we list in Table~\ref{tab:cosmos}. Their images and
unsharp masks are shown in Fig.~\ref{fig:cosmos}.

Based on their unsharp masks and ellipse fits (not shown), the objects
COSMOS J095922.76+024245.3 and COSMOS J095936.98+015107.6 clearly host
inner bars. Four of the other five galaxies probably contain nuclear
disks since the inner position angles match those of their outer
disks.  The remaining object is an S2B candidate. In this analysis,
the larger pixel scale of the COSMOS images as compared to GOODS has a
noticeable effect -- a drizzling process that would bring the COSMOS
pixel scale from the current $50$ mas down to $30$ mas as in GOODS
would result in a better discrimination of inner bars.
Unfortunately, no further data (such as colours and redshifts) are
publicly available for these galaxies; therefore their distances, bar
sizes, and stellar population properties remain unknown.


\section{Summary and Discussion}
\label{sec:discuss}

\begin{figure*}
{
\includegraphics[width=84mm]{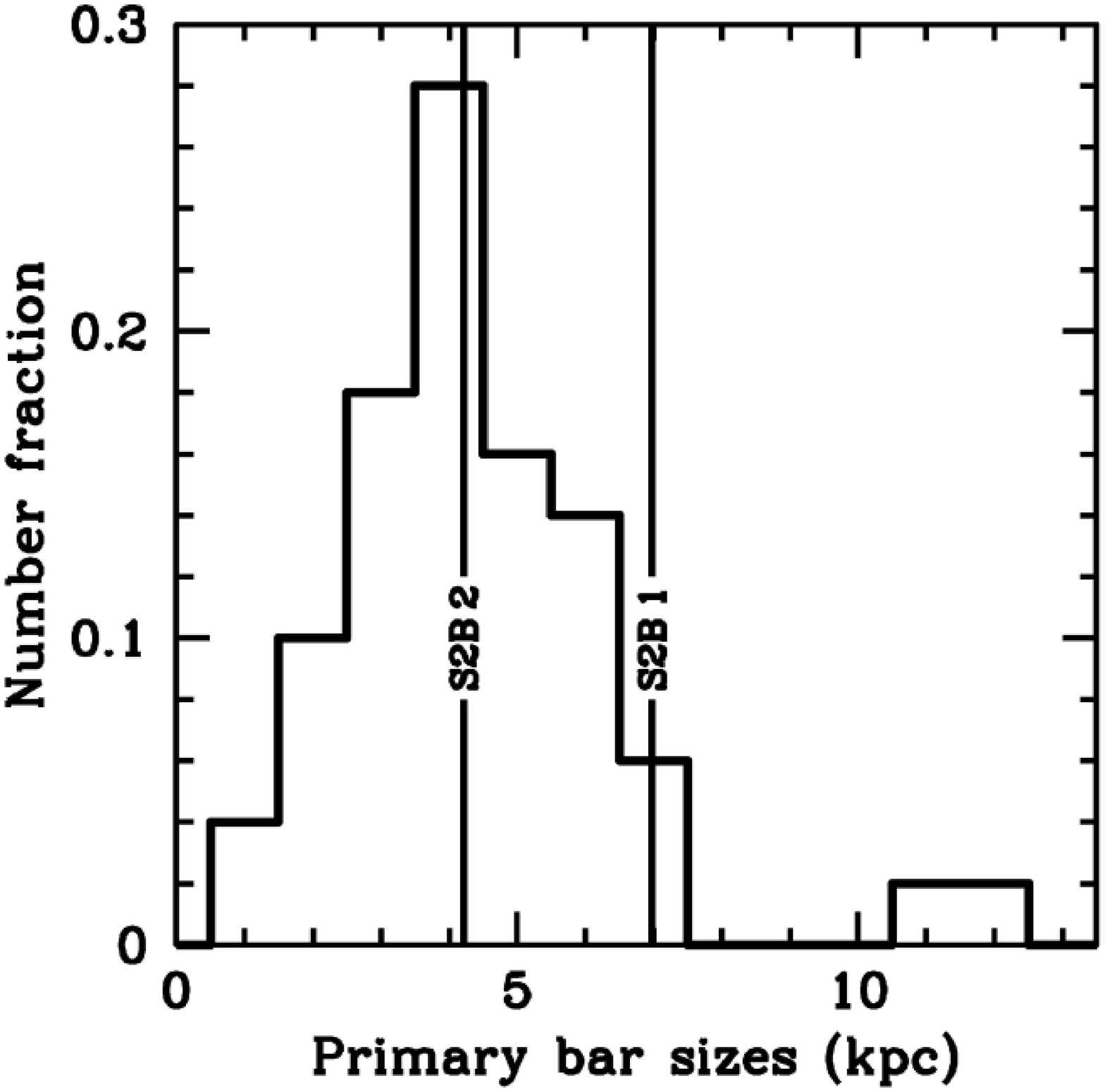}\\
\includegraphics[width=84mm]{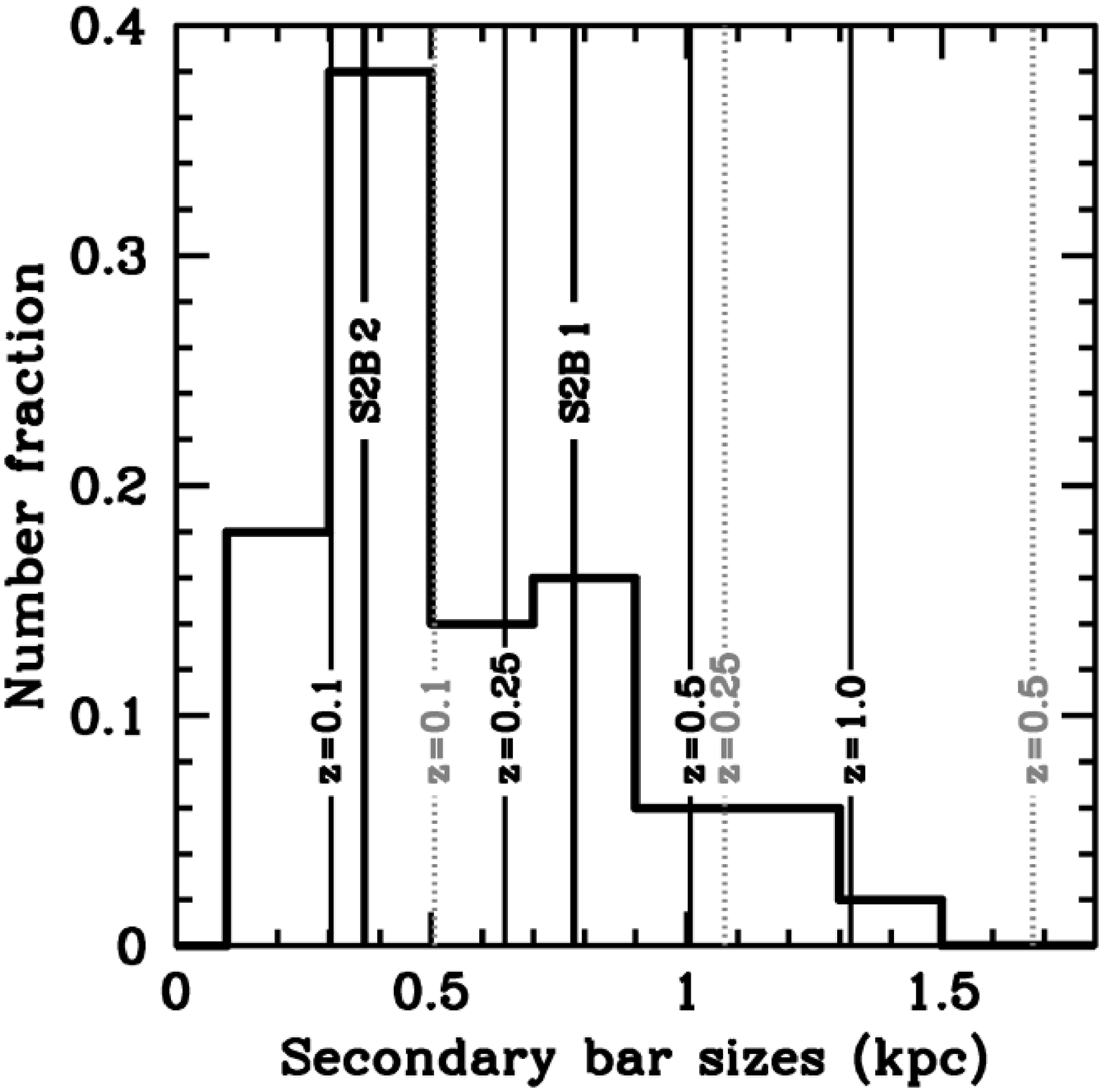}
}
\caption{{\bf Bar sizes.} Comparison of deprojected bar sizes
  ($a_{\epsilon}$) of the 50 nearby S2Bs from \citet{erw04} to our
  objects. \emph{Top:} Primary bar sizes. \emph{Bottom:} Secondary bar
  sizes.
Thick vertical lines denote the values for our double-bar detections (S2B\,1
  and S2B\,2). Vertical lines with redshift labels
give the minimum physical
  size of a 
  secondary bar that could still be identified at the respective
  redshift in the GOODS images or images with similar characteristics
  (pixel scale $30$ mas, $1$ PSF FWHM $= 0\farcs11$).
The dotted vertical lines assume a minimum angular size of the secondary bar's
  semi-major axis of $0\farcs28$ or $2.5$ PSF FWHM \citep{she03}, while the
  solid vertical lines assume a minimum angular size of $0\farcs17$ or $1.5$
  PSF FWHM, which is the semi-major axis of the smallest secondary bar we
  detect (S2B\,2).
}
\label{fig:sizes}
\end{figure*}

\begin{figure*}
\includegraphics[width=84mm]{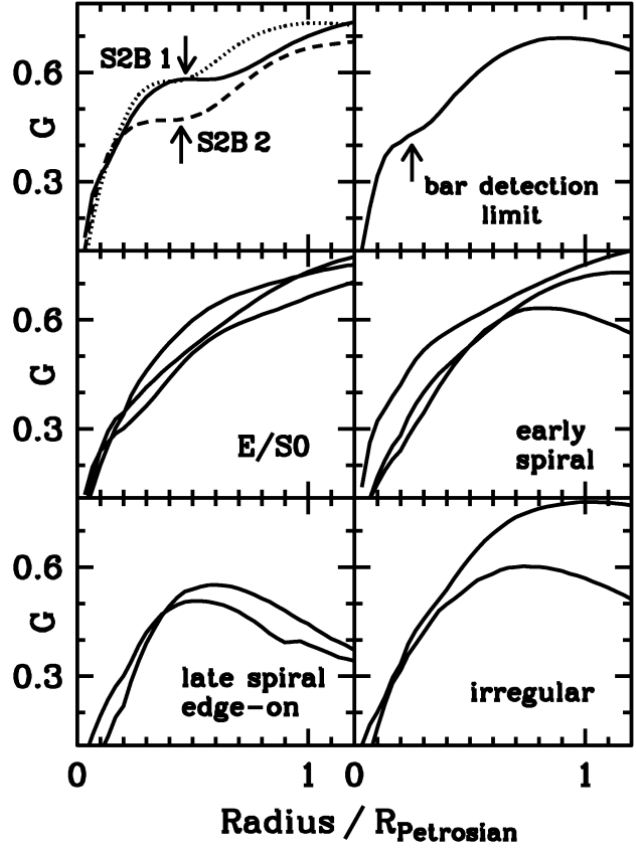}
\caption{{\bf Gini profiles.}
  The radial variation of the Gini
  coefficient ($G$) is shown for our two S2Bs (upper left panel), as
  well as for a sample of unbarred galaxies (middle and bottom panels)
  which are visually assigned to different classes as indicated in
  each panel. The dotted line in the upper left panel is for S2B\,1
  after artificially redshifting it to $z=0.3$ (see Sect.\
  \ref{sec:discuss}). Arrows indicate a plateau in the profile (see
  text for details).
  The upper right
  panel shows an early-type barred galaxy whose profile has a weak
  plateau just strong enough to be selected.
  The uppermost profile in the middle right panel
  belongs to an early-type spiral seen almost egde-on.
}
\label{fig:mopro}
\end{figure*}

\begin{figure*}
{
\includegraphics[width=60mm]{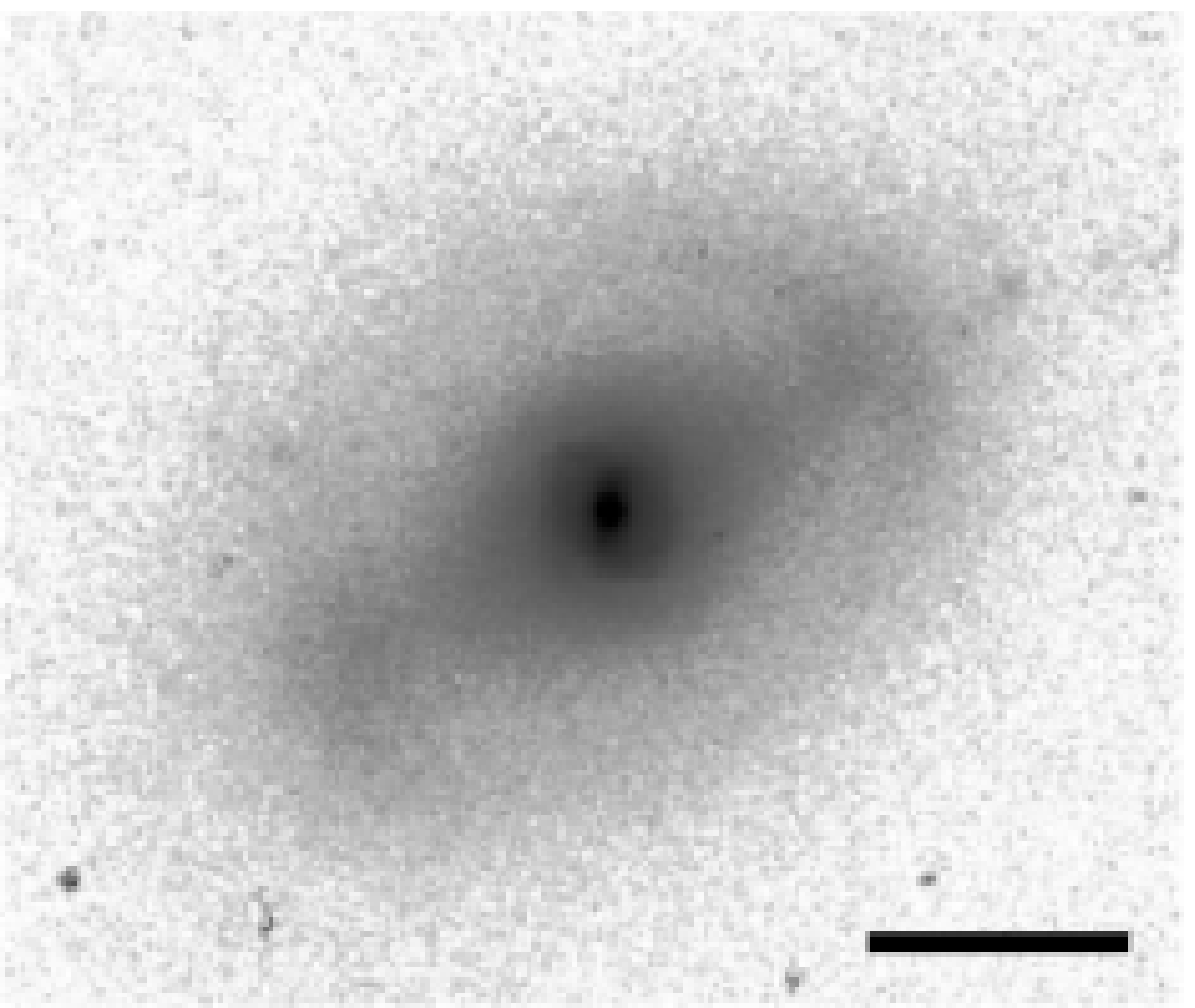}\includegraphics[width=60mm]{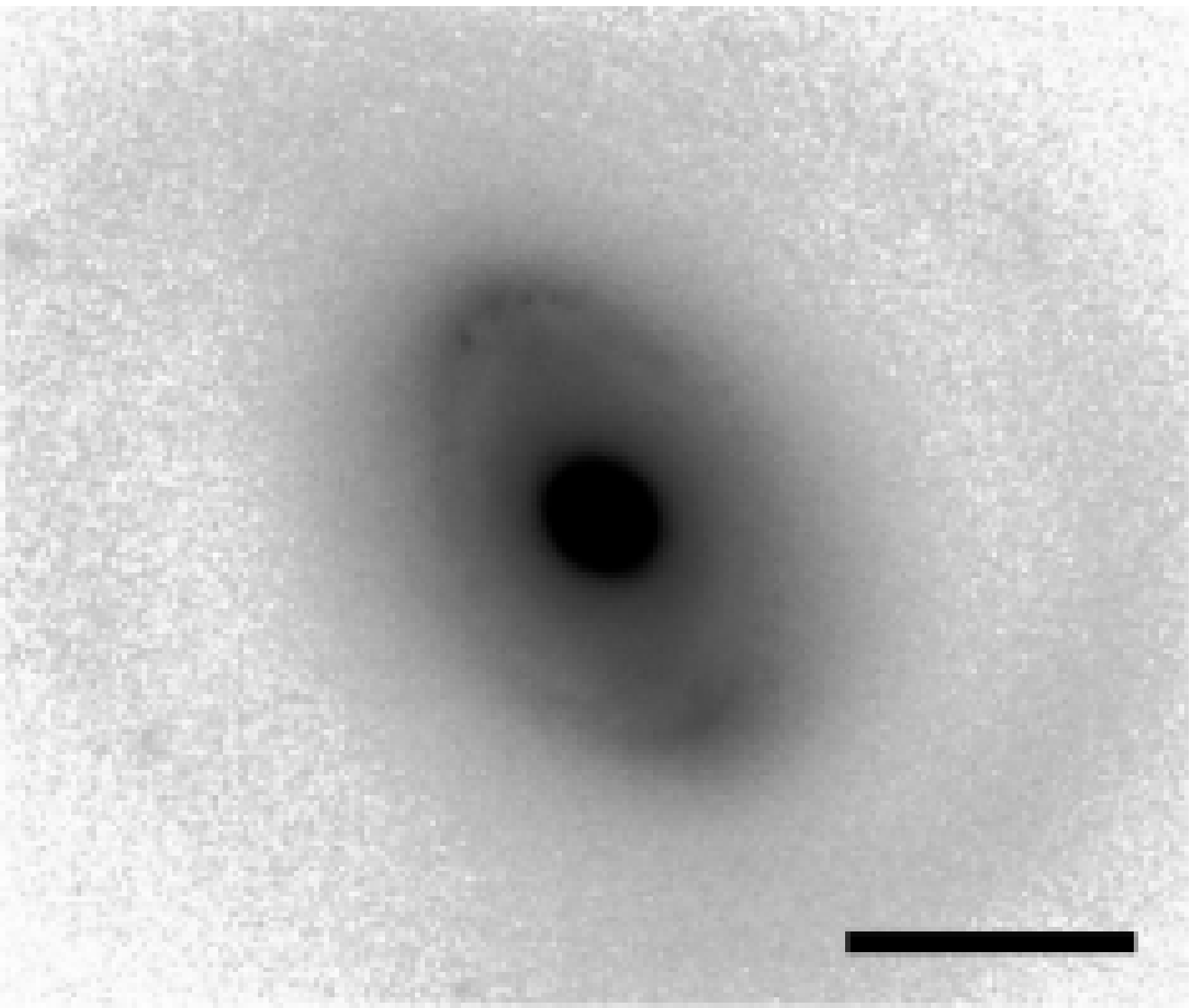} 
\includegraphics[width=60mm]{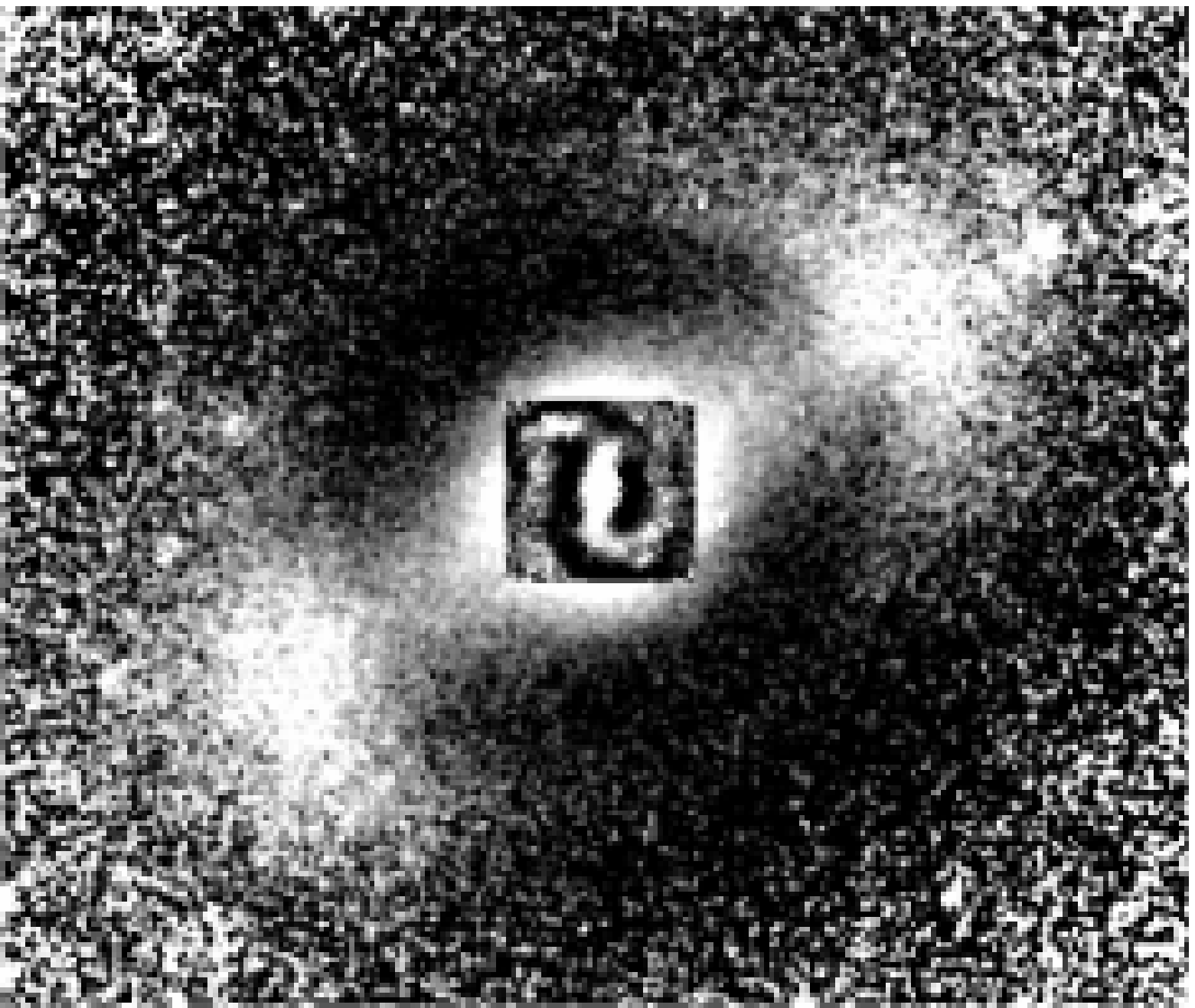}\includegraphics[width=60mm]{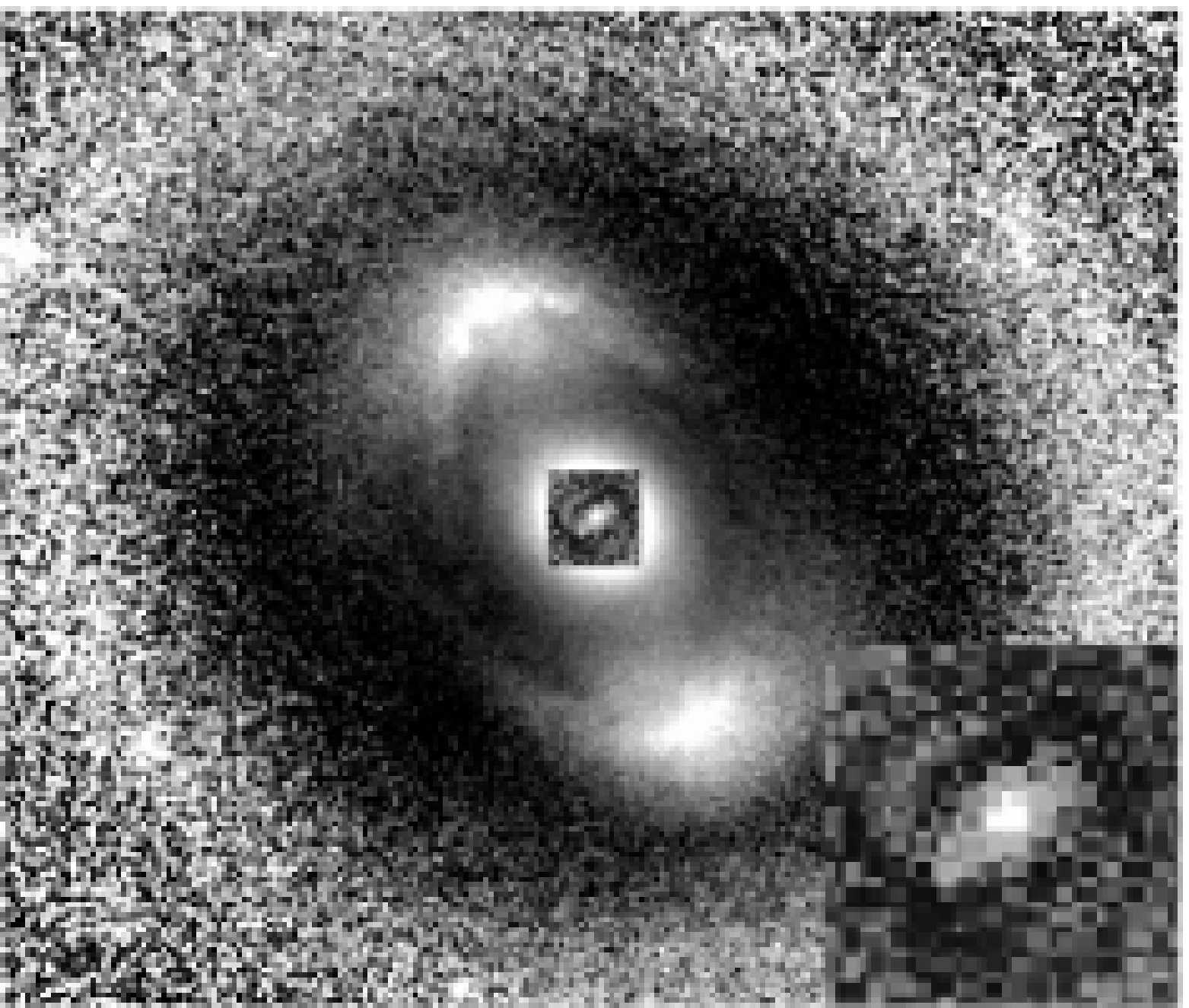}
\includegraphics[width=60mm]{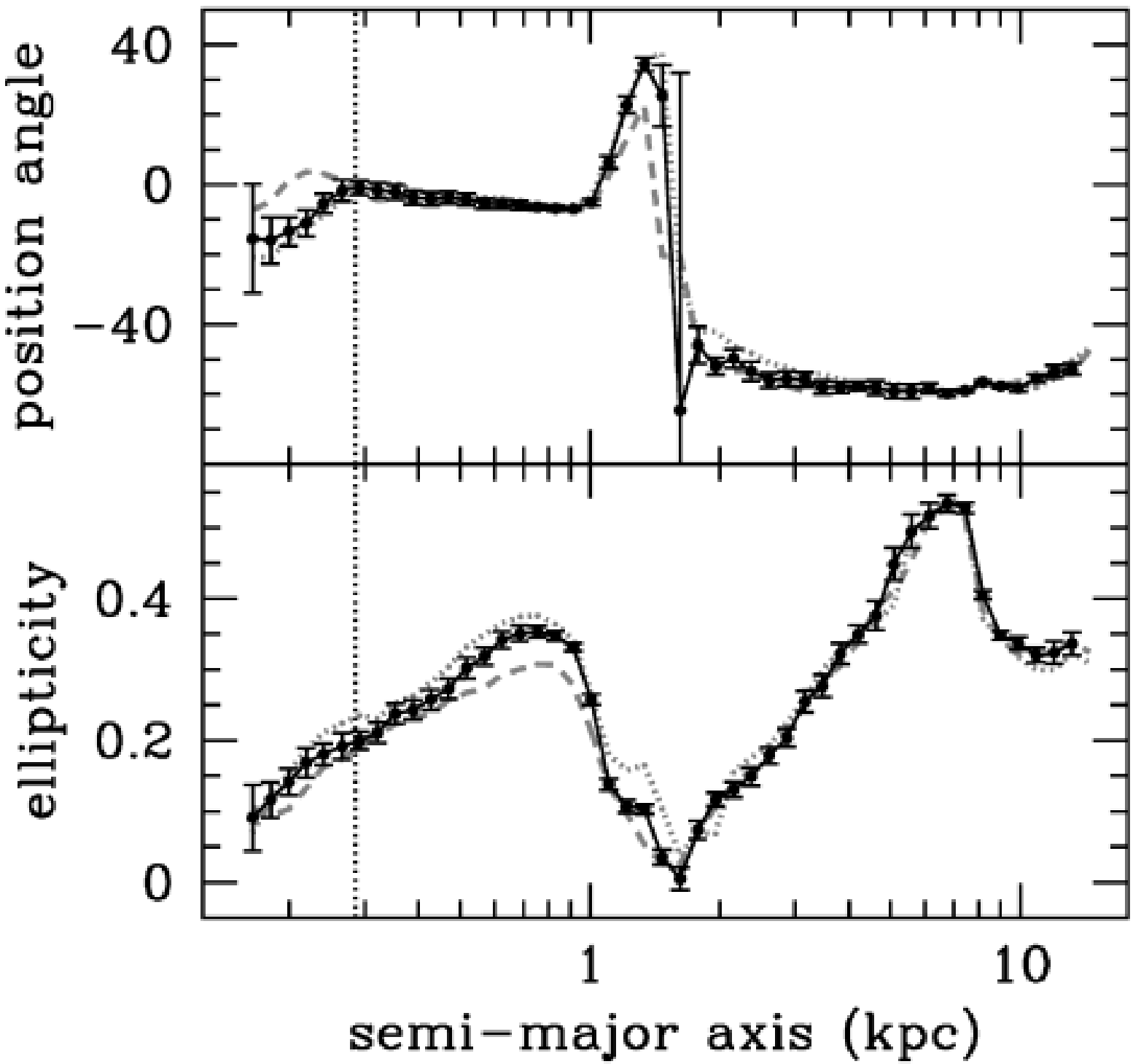}\includegraphics[width=60mm]{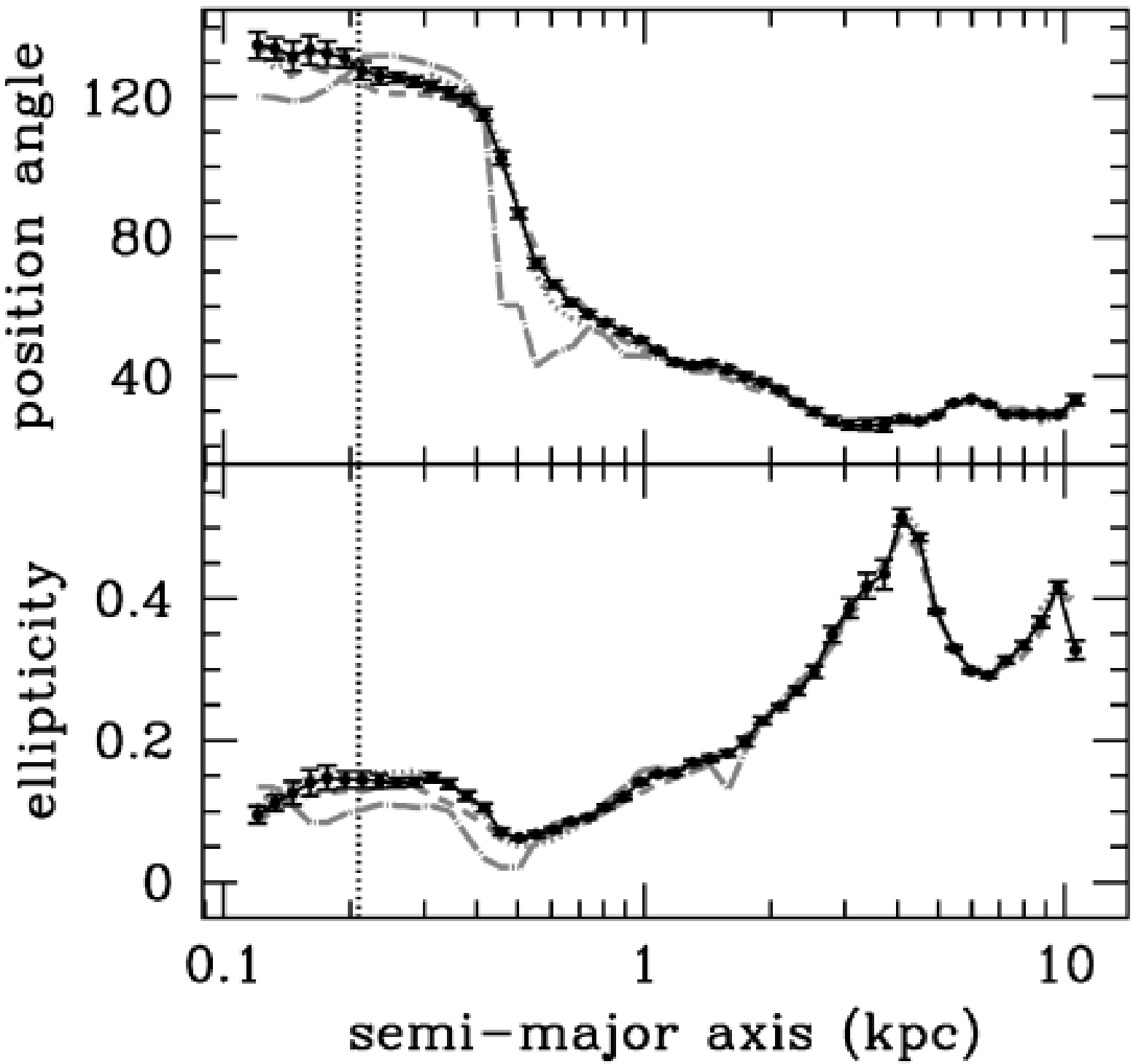}
}
\caption{{\bf Images, unsharp masks \& ellipse fits.} For S2B\,1 (left
  panels) and S2B\,2 we present $i$-band images (top panels), unsharp
  masks (middle panels), and ellipse fits (lower panels). In the
  images, north is up and east is to the left; position angle is
  measured east from north. The black scale bar corresponds to a
  physical length of $5\,\rm{kpc}$. Unsharp masks have been computed
  by dividing the original $i$-band image by a Gaussian-convolved
  image. Two unsharp masks are superposed: one using a large Gaussian
  $\sigma$ ($\sigma=40\,\rm{pix}$ for S2B\,1 and $\sigma=30\,\rm{pix}$
  for S2B\,2), and a second one using a small value
  ($\sigma=4\,\rm{pix}$ for S2B\,1 and $\sigma=1.5\,\rm{pix}$ for
  S2B\,2) which is shown as inset in the image center and 
is enlarged in the lower right corner for S2B\,2.
Ellipse fits (using the IRAF task ELLIPSE)
  are shown in black for the $i$-band images, with errors as given by
  ELLIPSE. The other bands are only shown as lines: dash-dotted for
  $B$-band, dotted for $V$-band, and dashed for $z$-band.
  The dotted vertical line indicates the value of 1 PSF FWHM
  ($0\farcs11$).}
\label{fig:images}
\end{figure*}

\begin{figure*}
\begin{center}
\includegraphics[width=60mm]{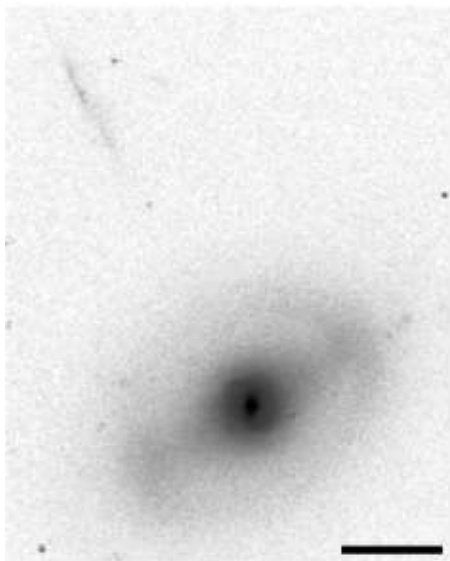}
\end{center}
\caption{{\bf A jet?} $i$-band image of S2B\,1 and the jet-like
  object pointing back to its center. The black scale bar corresponds
  to a physical length of $5\,\rm{kpc}$.}
\label{fig:barryouter}
\end{figure*}

\begin{figure*}
\includegraphics[height=55mm]{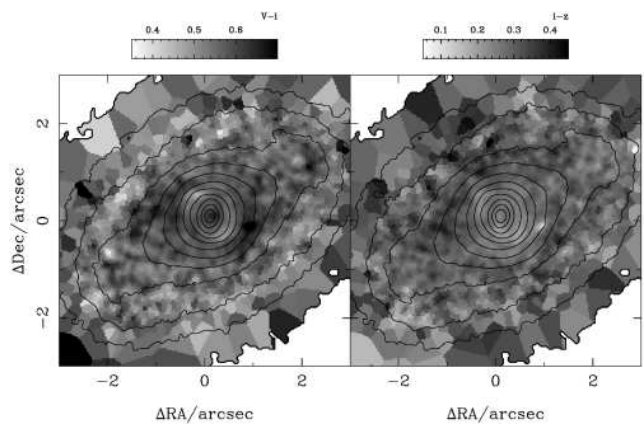}
\caption{{\bf Colour maps of S2B\,1.} An adaptive binning based on the
Optimal Voronoi Tessellation of \citet{cap03} has been applied.  A
target SNR $=20$ was chosen in the flux ratio between the $V$ and the
$i$ passbands. The contours give the $i$-band surface brightness from
18~mag/arcsec$^2$ (inner contour) in steps of 0.5~mag.}
\label{fig:BC}
\end{figure*}

\begin{figure*}
\includegraphics[height=60mm]{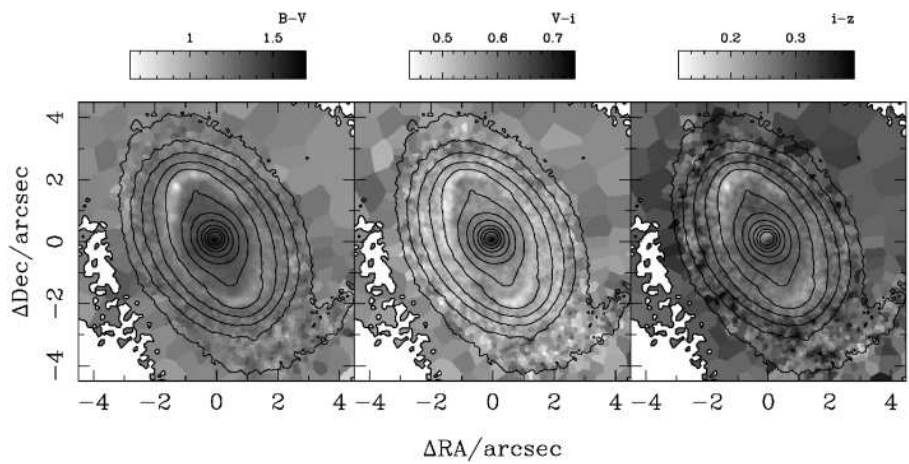}
\caption{
{\bf Colour maps of S2B\,2.} Similar to Figure~\ref{fig:BC}.}
\label{fig:MC}
\end{figure*}

\begin{figure*}
\includegraphics[width=84mm]{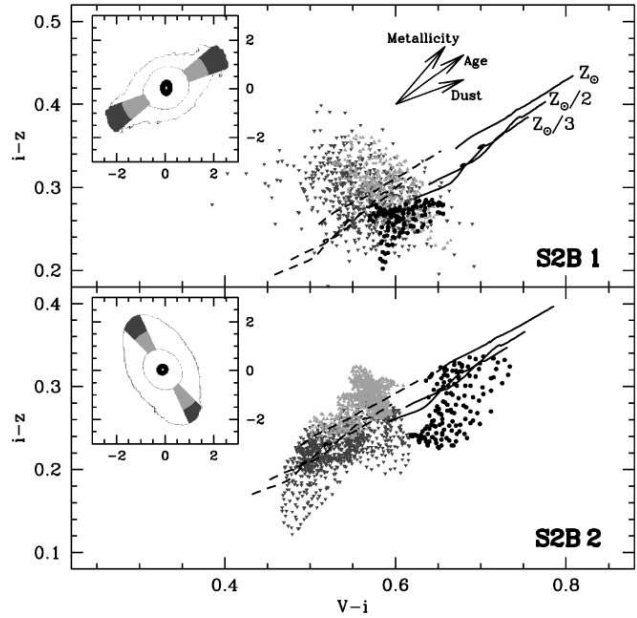}
\caption{ {\bf Colour-Colour diagrams.} Colours are given separately for
three regions in each galaxy (see $6\arcsec \times 6\arcsec$ inset).
Pixels in the outer bar are selected from a $20\degrees$ wedge aligned
with the bar and chosen to avoid dust lanes. Filled triangles
represent regions in the outer bar at radial distances $r<1\farcs8$
(light gray, pointing upward) or $r>1\farcs8$ (dark gray, pointing
downward).  The filled dots are the colours at the position of
the inner bar, chosen between 17.5 and 18.5~mag/arcsec$^2$ (S2B\,1) and
between 17 and 18~mag/arcsec$^2$ (S2B\,2). We exclude the very central
region which could be
significantly reddened by dust. The dashed and solid lines track the
colours of simple stellar populations with $E(B-V)=0$ and $0.2$,
respectively. Each line is an age sequence from 2 to 10 Gyr. See text
for more details.}
\label{fig:ccd}
\end{figure*}

\begin{figure*}
{
\includegraphics[width=31mm]{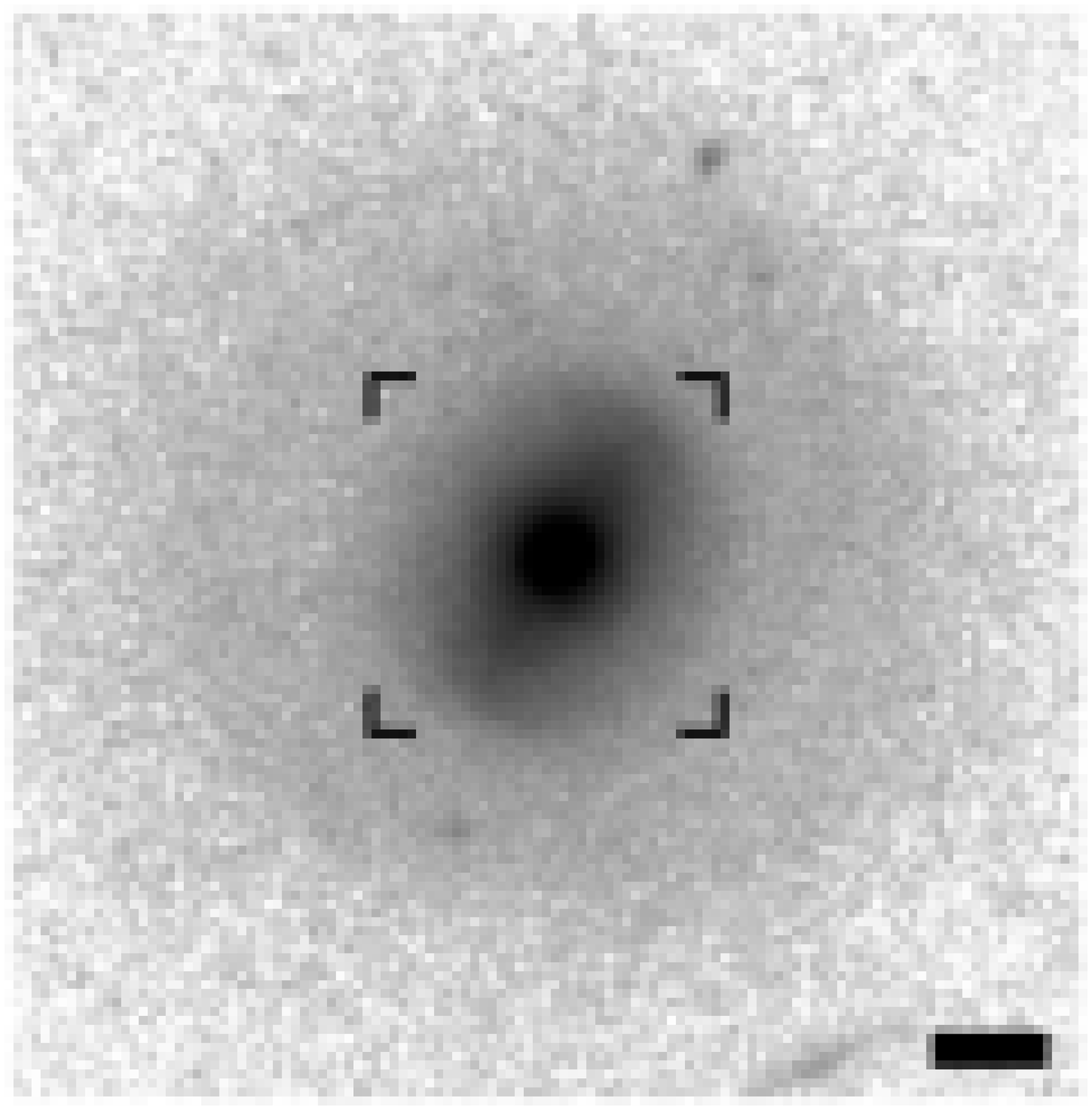}\includegraphics[width=31mm]{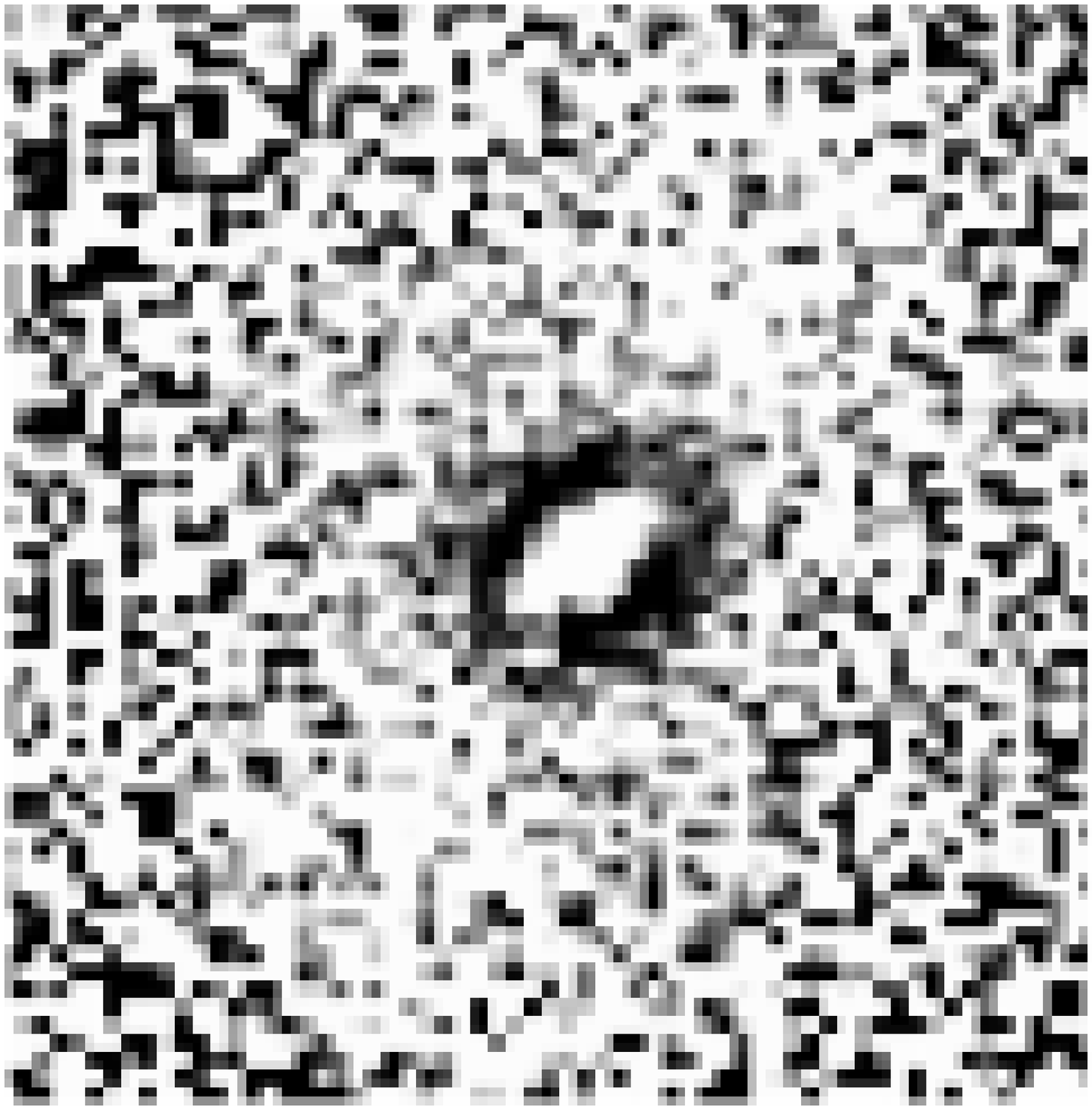}\includegraphics[width=31mm]{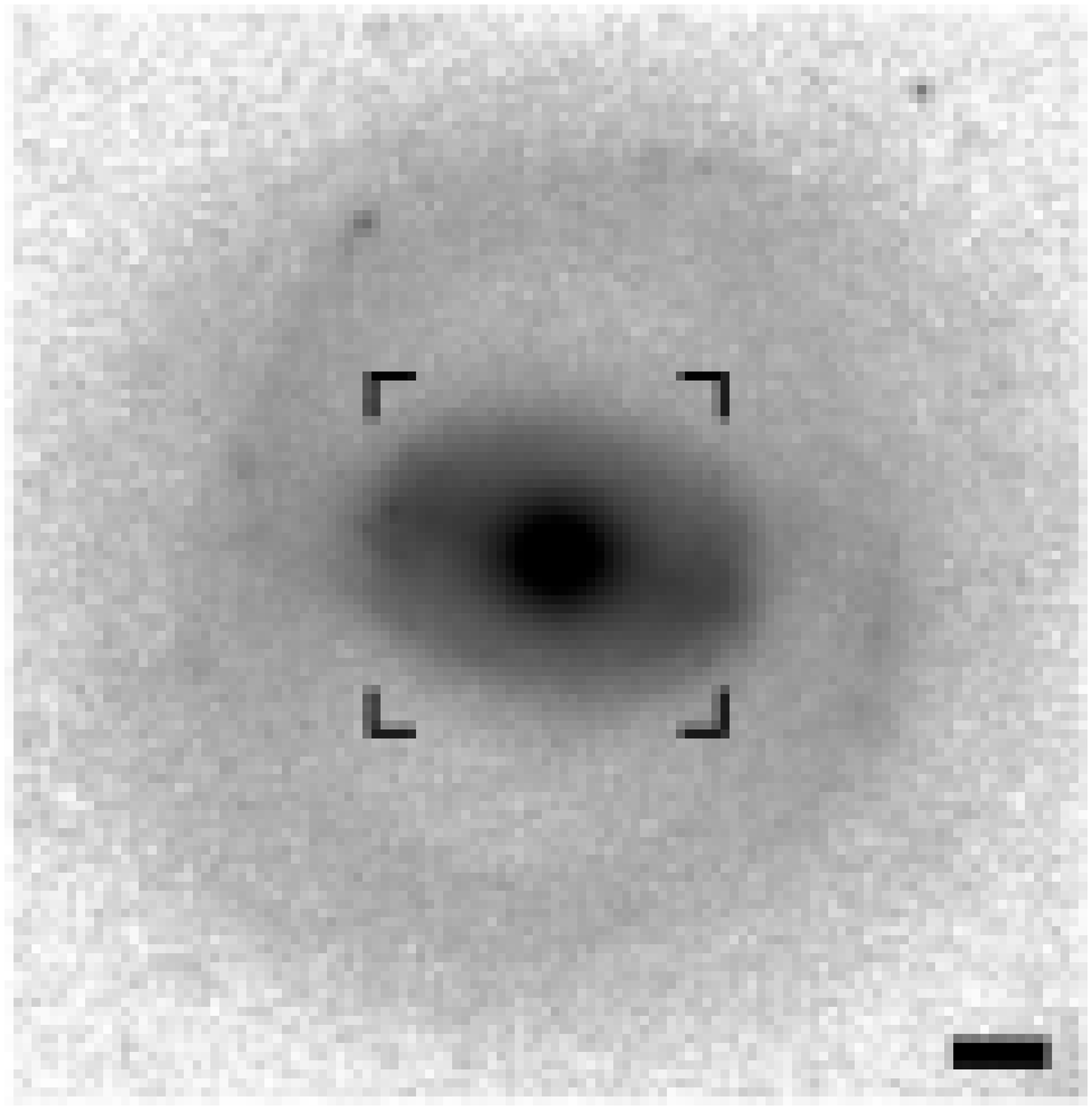}\includegraphics[width=31mm]{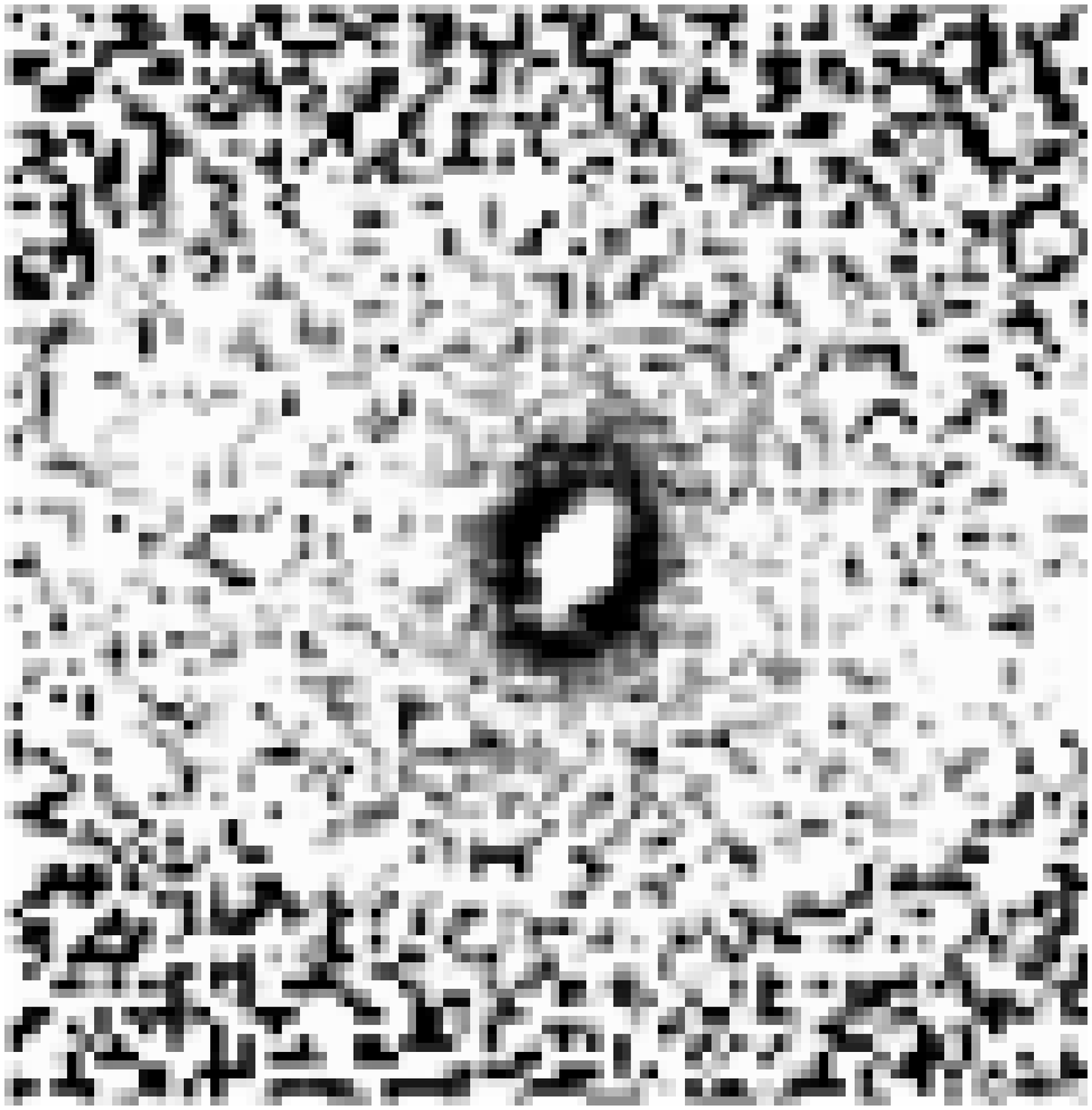}
\includegraphics[width=31mm]{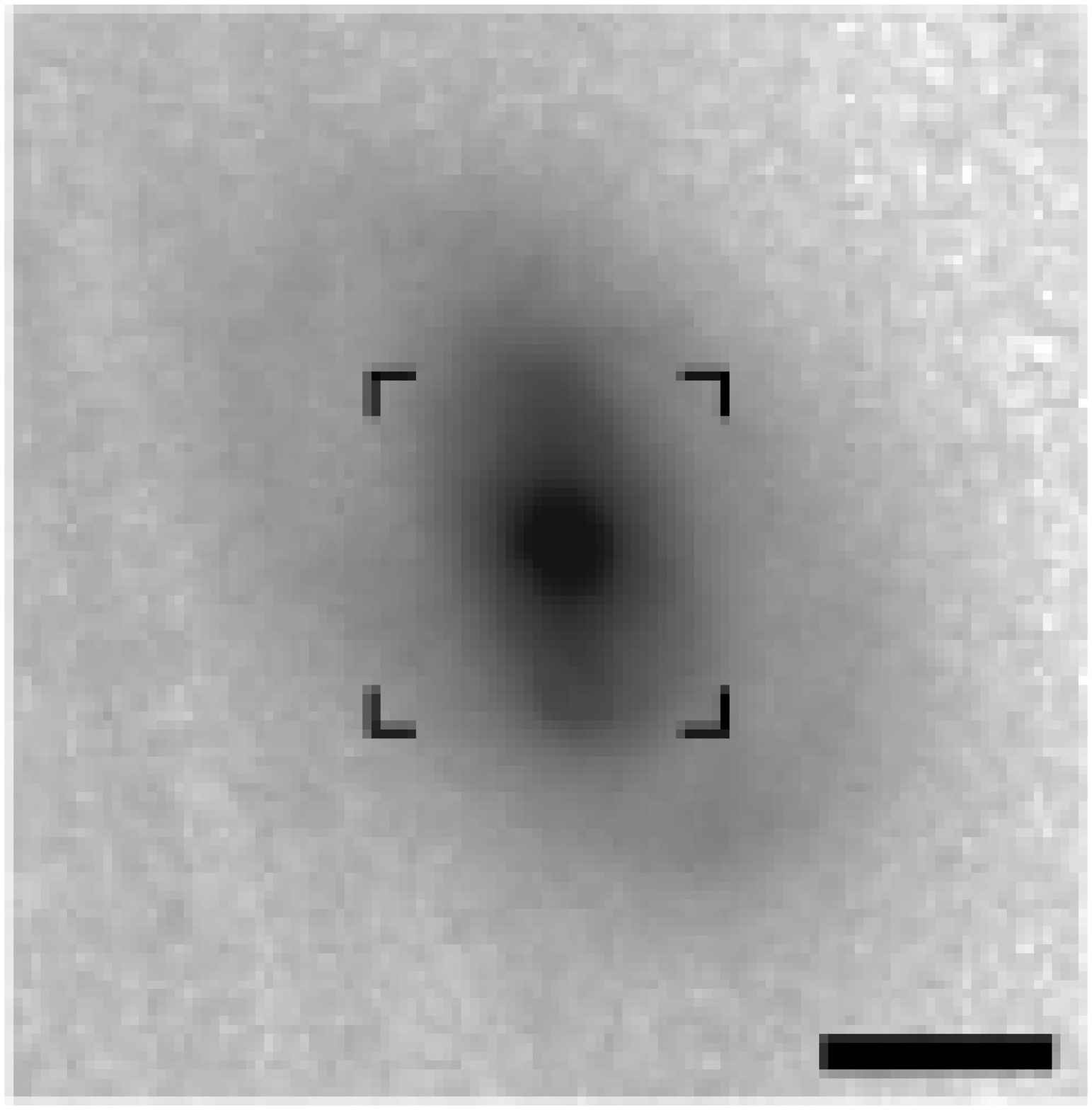}\includegraphics[width=31mm]{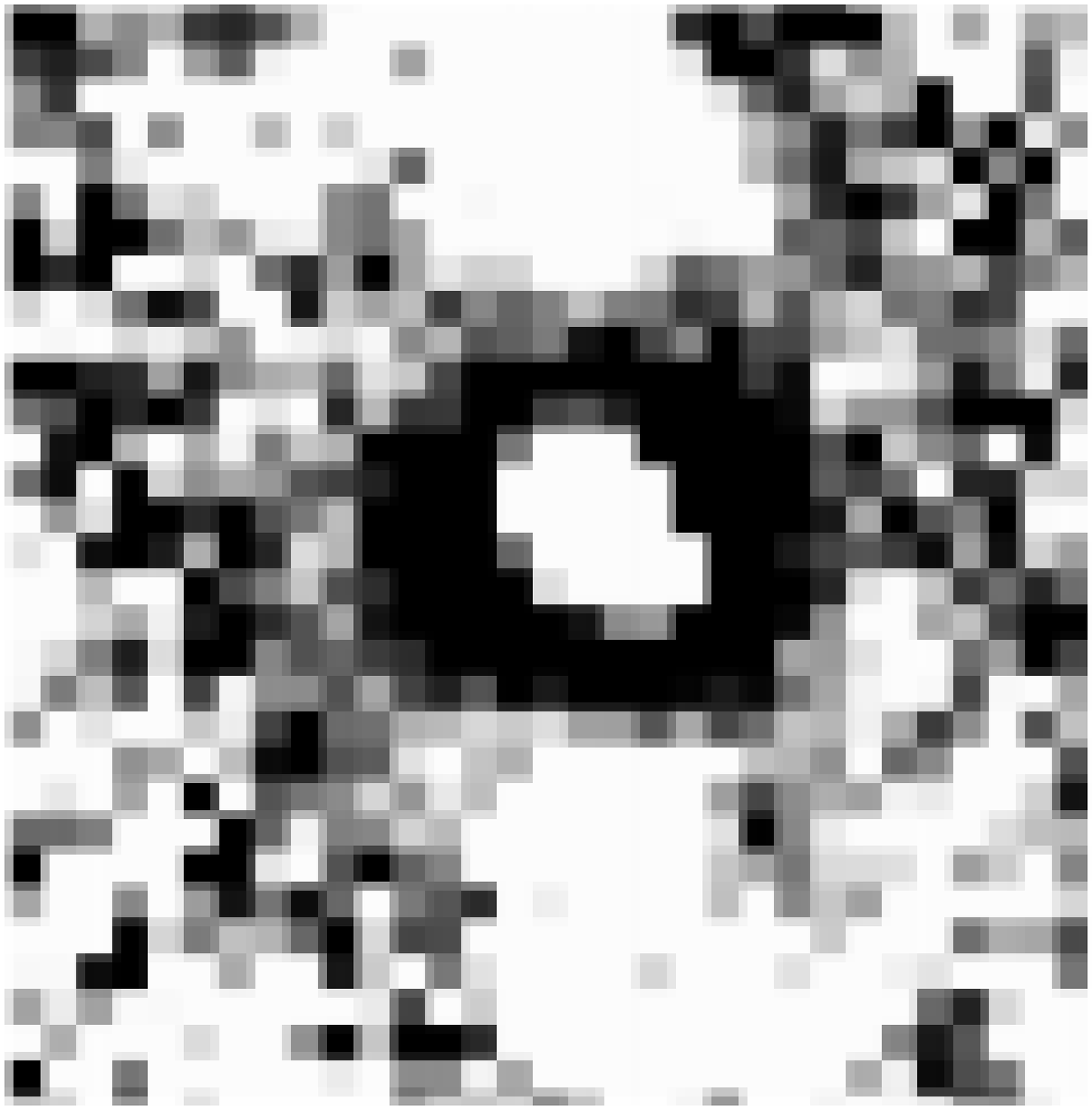}\includegraphics[width=31mm]{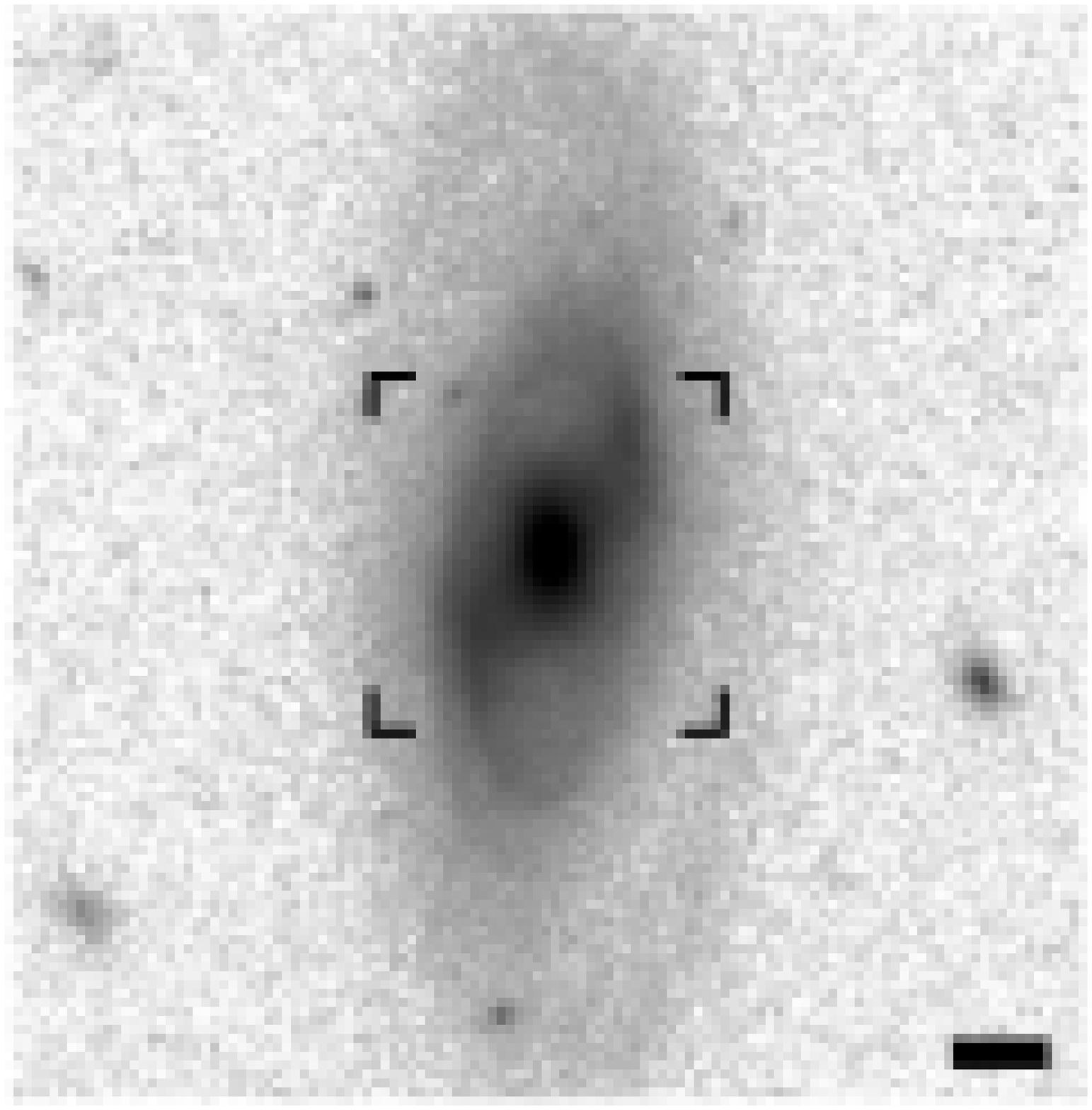}\includegraphics[width=31mm]{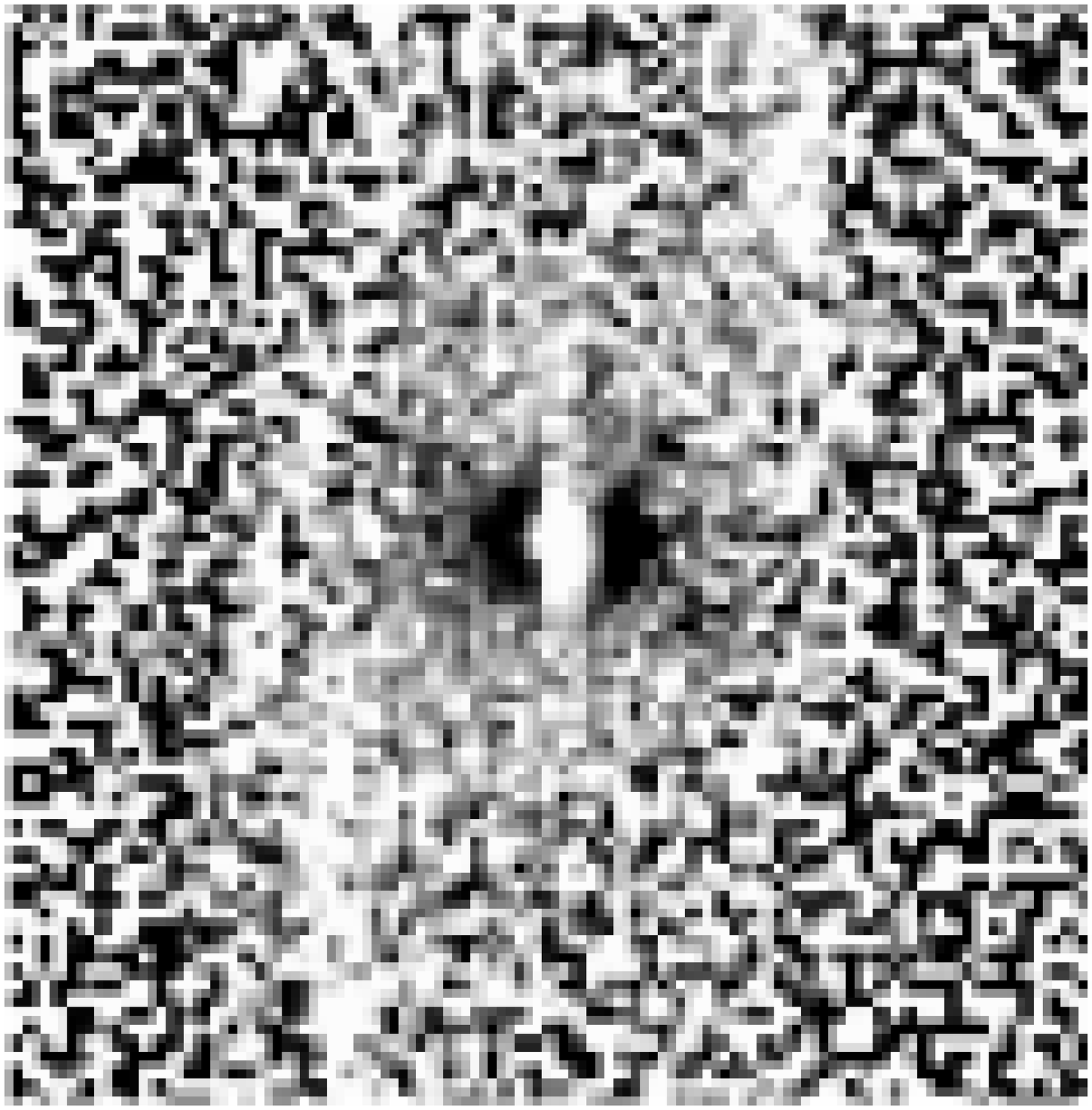}
\includegraphics[width=31mm]{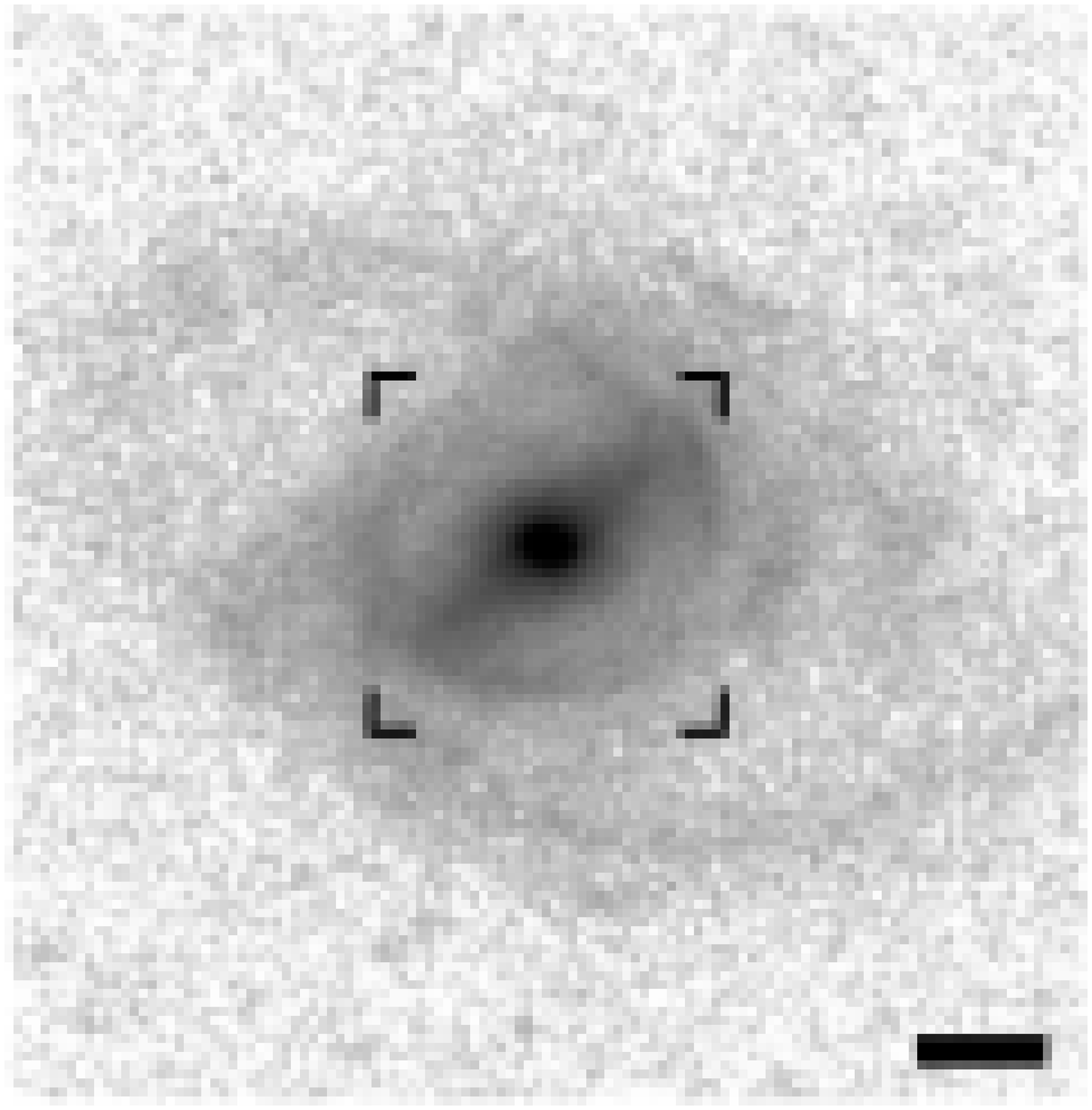}\includegraphics[width=31mm]{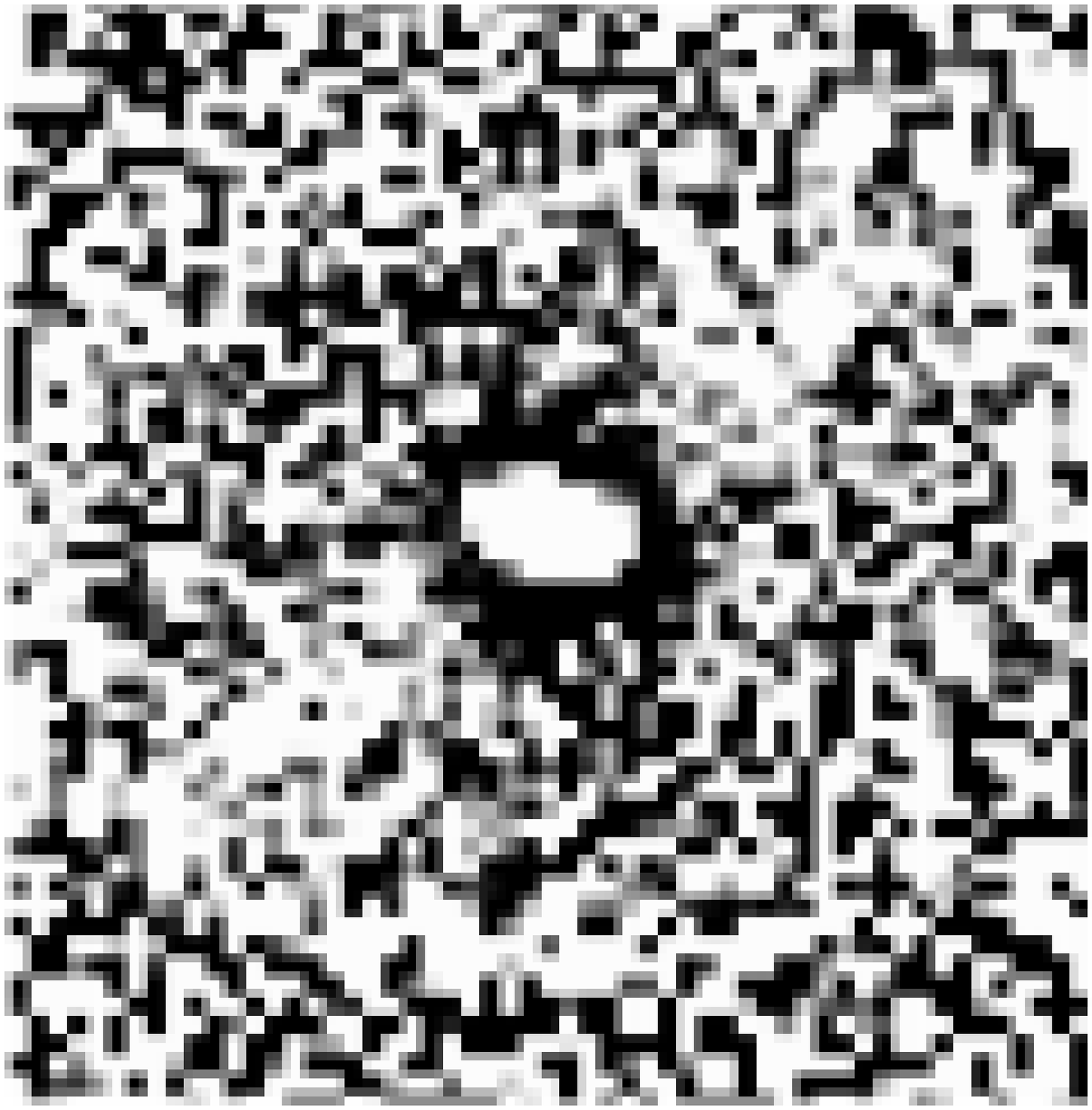}\includegraphics[width=31mm]{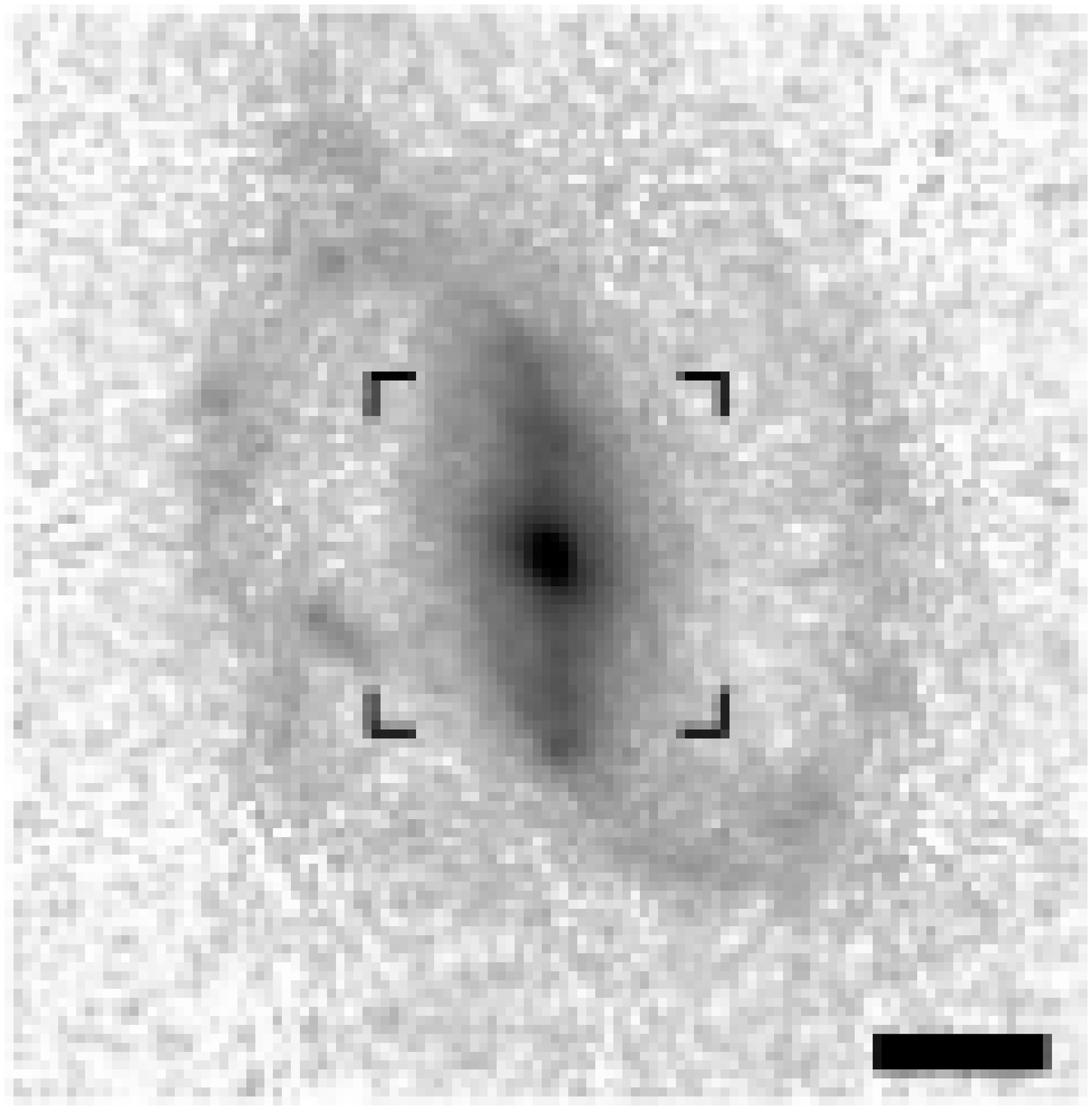}\includegraphics[width=31mm]{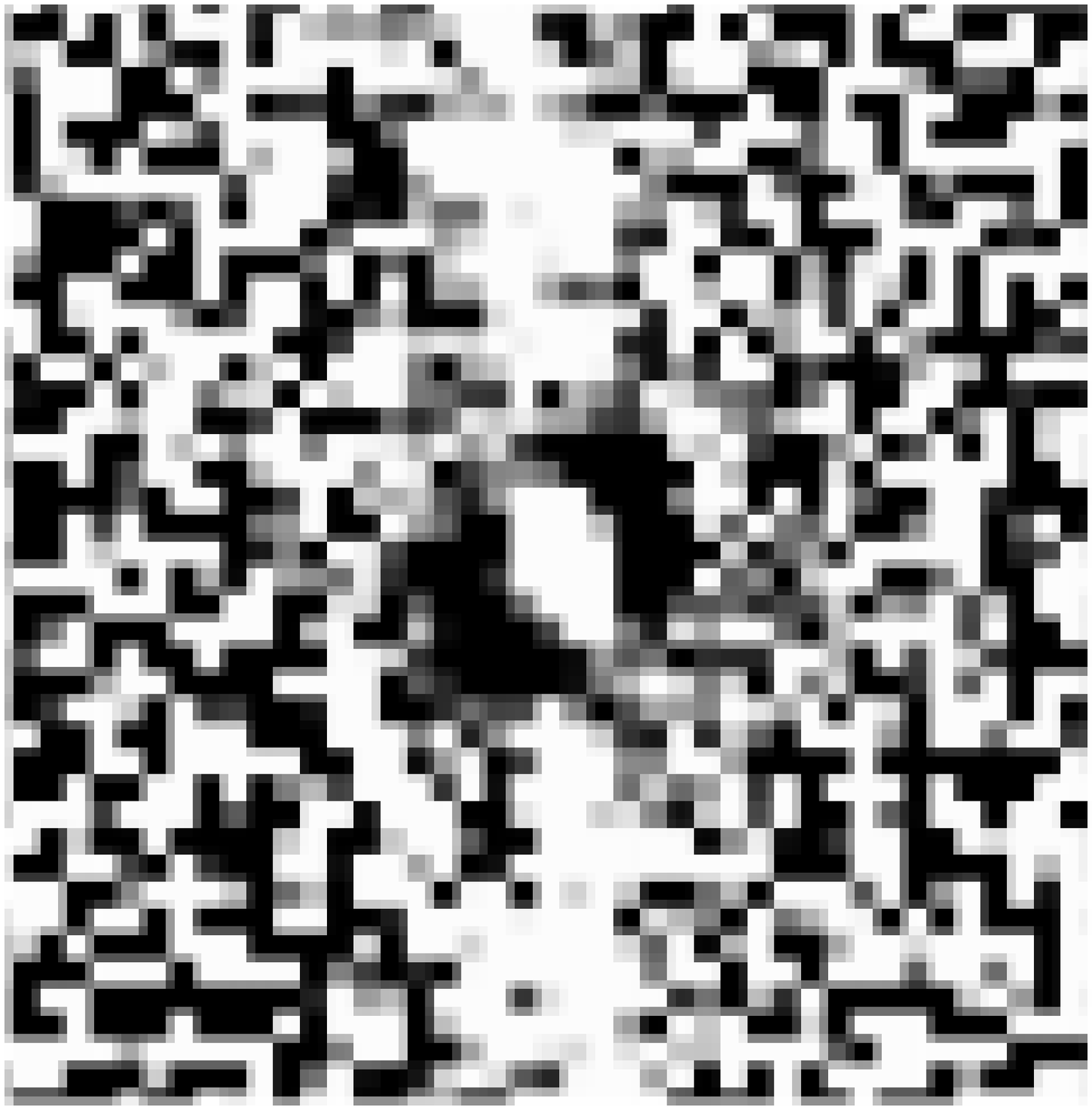}
\includegraphics[width=31mm]{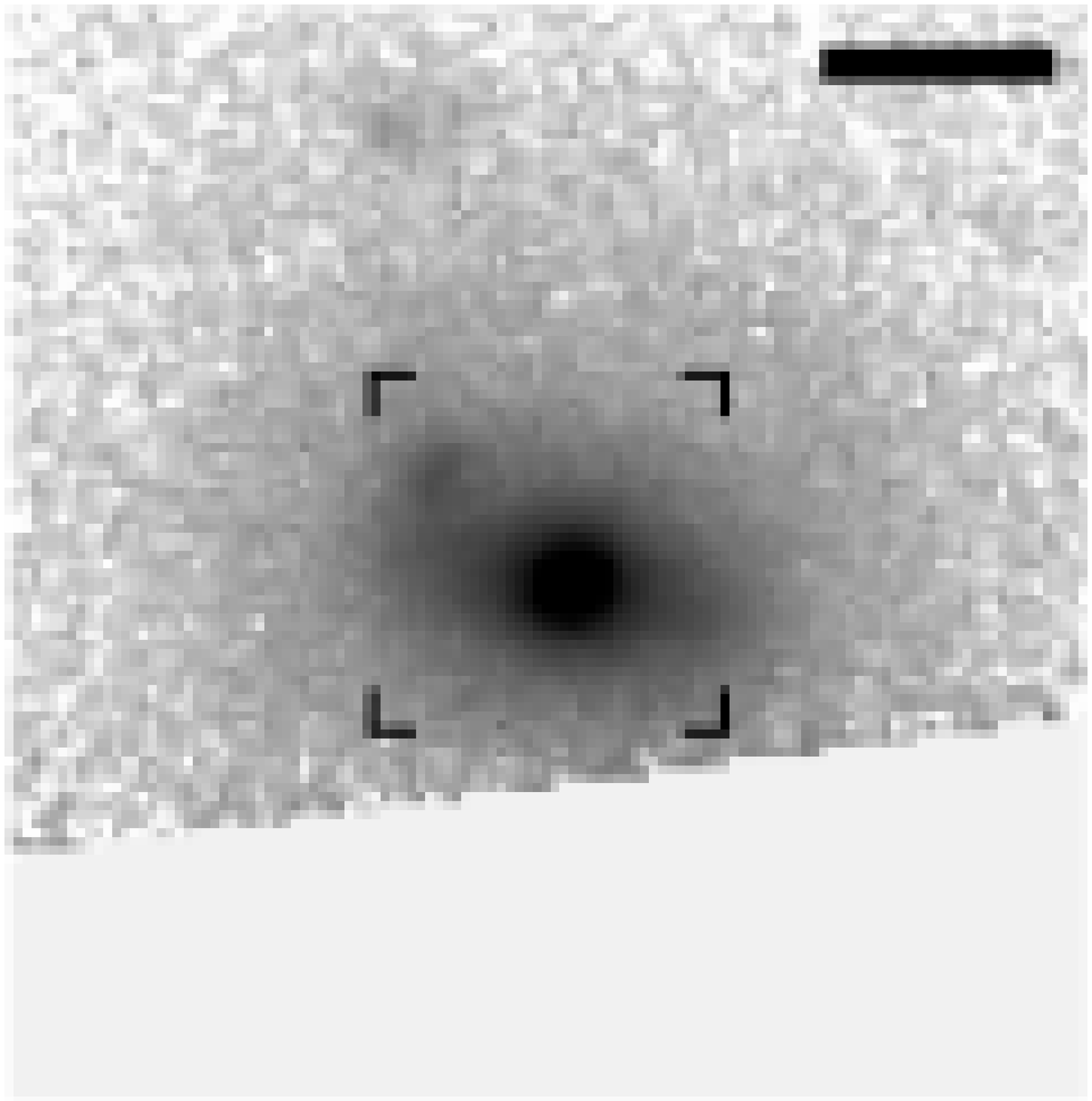}\includegraphics[width=31mm]{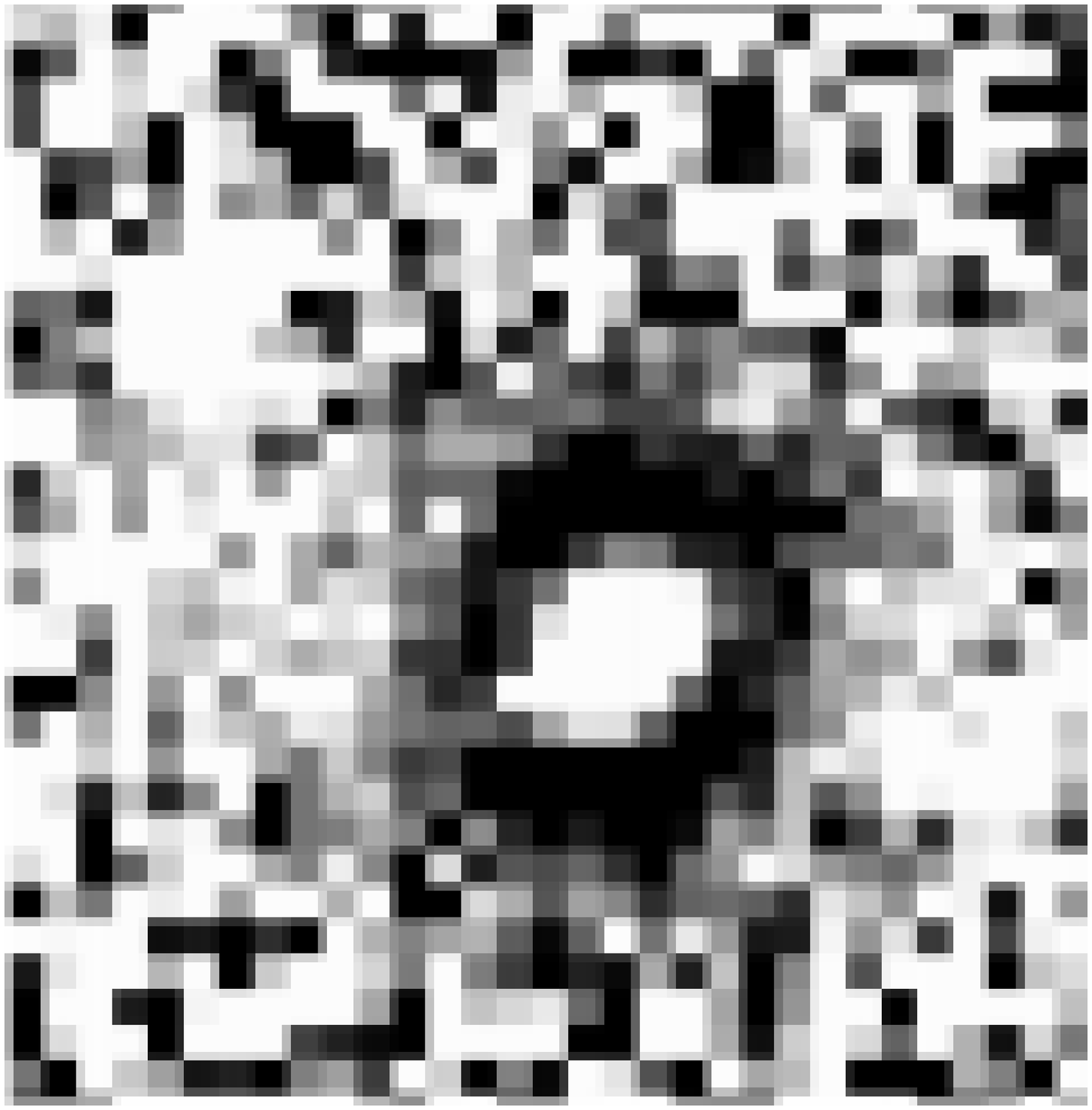}\hspace{61mm}
}
\caption{{\bf S2Bs and S2B candidates in COSMOS.} Objects are shown in
  the same order as in Table~\ref{tab:cosmos}, i.e.\ the top two
  objects are classified as S2B, the middle four objects as possible
  S2B or SB with nuclear disk, and the lowermost object as possible
  S2B. For each object, the $i$-image is shown
  on the left and the unsharp mask created with a Gaussian of $\sigma = 1.5$ pixels on the
  right. North is up and east is
  to the left. The black scale bar corresponds to $1\arcsec$. The masks
  only cover the central area of each object, as indicated by the box edges in each image.}
\label{fig:cosmos}
\end{figure*}

To date, all observational studies of S2Bs have been undertaken in the
nearby universe.  In this paper we presented the two most distant S2Bs
currently known, at $z=0.103$ and $z=0.148$,
corresponding to a look-back time of 1.3 and 1.9\,Gyr,
respectively.  Although our S2B sample contains only two objects, we
can perform a consistency check with the number of S2Bs that would be
expected given the numbers for nearby objects. Our GOODS sample of
early-type barred galaxies consists of 17 objects that were selected
by our Gini profile method, plus 11 barred galaxies that were missed
by the method (see Sect.\ \ref{sec:morph}). Adopting an S2B fraction
of 30\% \citep[e.g.][]{erw02} we would thus expect
8.4 S2Bs to be within our sample. With a given distribution of
redshifts, the bottom panel of Fig.\ \ref{fig:sizes} then leads to an
estimate of the number of S2Bs whose inner bar would be large enough
to be identified. Redshifts are available for 12 SBs in GOODS/CDFS
from the COMBO-17 survey \citep{combo17} and the VIMOS VLT Deep Survey
\citep[VVDS,][]{vvds}, with 6 objects beyond $z=0.5$. When we adopt
this distribution to be representative for all 28 SB galaxies, we
would expect to detect $2.4\pm1.5$ S2Bs, in excellent accordance with
our findings. While this suggests no significant evolution of the S2B
fraction, clearly a larger sample of S2Bs at intermediate redshifts is
necessary to address the statistics properly.

\begin{figure*}
\includegraphics[width=84mm]{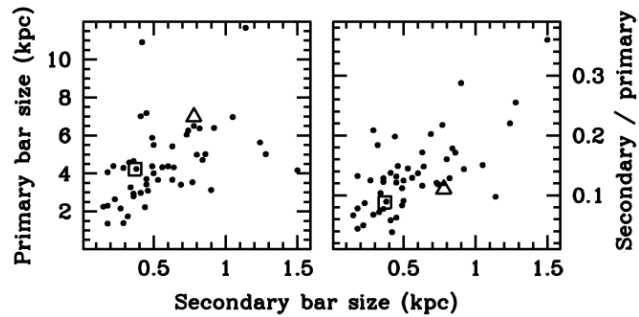}
\caption{{\bf Size relations of inner and outer bars.}
  Comparison of deprojected sizes
  ($a_{\epsilon}$) of inner and outer bars for the 50 nearby S2Bs from
  \citet[filled dots]{erw04} and for S2B\,1 (open triangle) and
  S2B\,2 (open square). \emph{Left:} Primary versus
  secondary bar size. \emph{Right:} Ratio of secondary-to-primary
  size versus secondary bar size.
  }
\label{fig:sizes2}
\end{figure*}

Any sample of intermediate-redshift S2Bs would of course be biased
towards large secondary bars, and also towards large primary bars --
the latter being due to both the selection of barred galaxy candidates
and due to a correlation of the sizes of inner and outer bar. Despite
this inevitable bias, it would be possible to compare the
\emph{relations} between the sizes of primary and secondary bar with
those of the local sample: the data from \citet{erw04} show a clear
correlation of primary and secondary bar size (left panel of Fig.\
\ref{fig:sizes2}). Moreover, the 
secondary-to-primary ratio is not constant but it increases with
increasing secondary bar size (right panel of Fig.\
\ref{fig:sizes2}). If an intermediate-redshift sample 
showed an offset with respect to the local relations, this would hint
at a decoupled evolution of sizes, possibly related to the loss of
angular momentum. While S2B\,2 (open square in Fig.\ \ref{fig:sizes2})
falls onto the above relations, the primary bar of S2B\,1 (open
triangle) is very large compared with its secondary. Still, S2B\,1
falls within the scatter of the relations. 

In addition to the structural parameters, we presented a colour
analysis of the two S2Bs.  Unfortunately, it is not possible from the
available data alone to precisely trace back the star formation
history of an individual galaxy (and of its bars), especially given
the possible effect of dust and the well-known age-metallicity
degeneracy. Nevertheless, we show that population differences between
the outer bar and the inner bar region can be quantified,
demonstrating the feasibility of stellar population analyses of S2Bs
at intermediate redshifts.  It is not clear whether the evolution of
the primary bar and the secondary's stellar content are linked in some
way. For example, the secondary bar could be formed from stars born
from gas swept by the primary bar. An intermediate-redshift S2B sample
of significant size would be able to address this question.  If the
amount of dust were small, the colours of S2B\,1 could be interpreted
as the inner bar having lower metallicity and larger stellar ages
(Sect.\ \ref{sec:barrycol}), making the above scenario implausible for this
galaxy.
The
colours of S2B 2 are clearly more strongly affected by dust and little
can be said about the relative ages of the primary and secondary bars.

\begin{figure*}
{
\includegraphics[width=50mm]{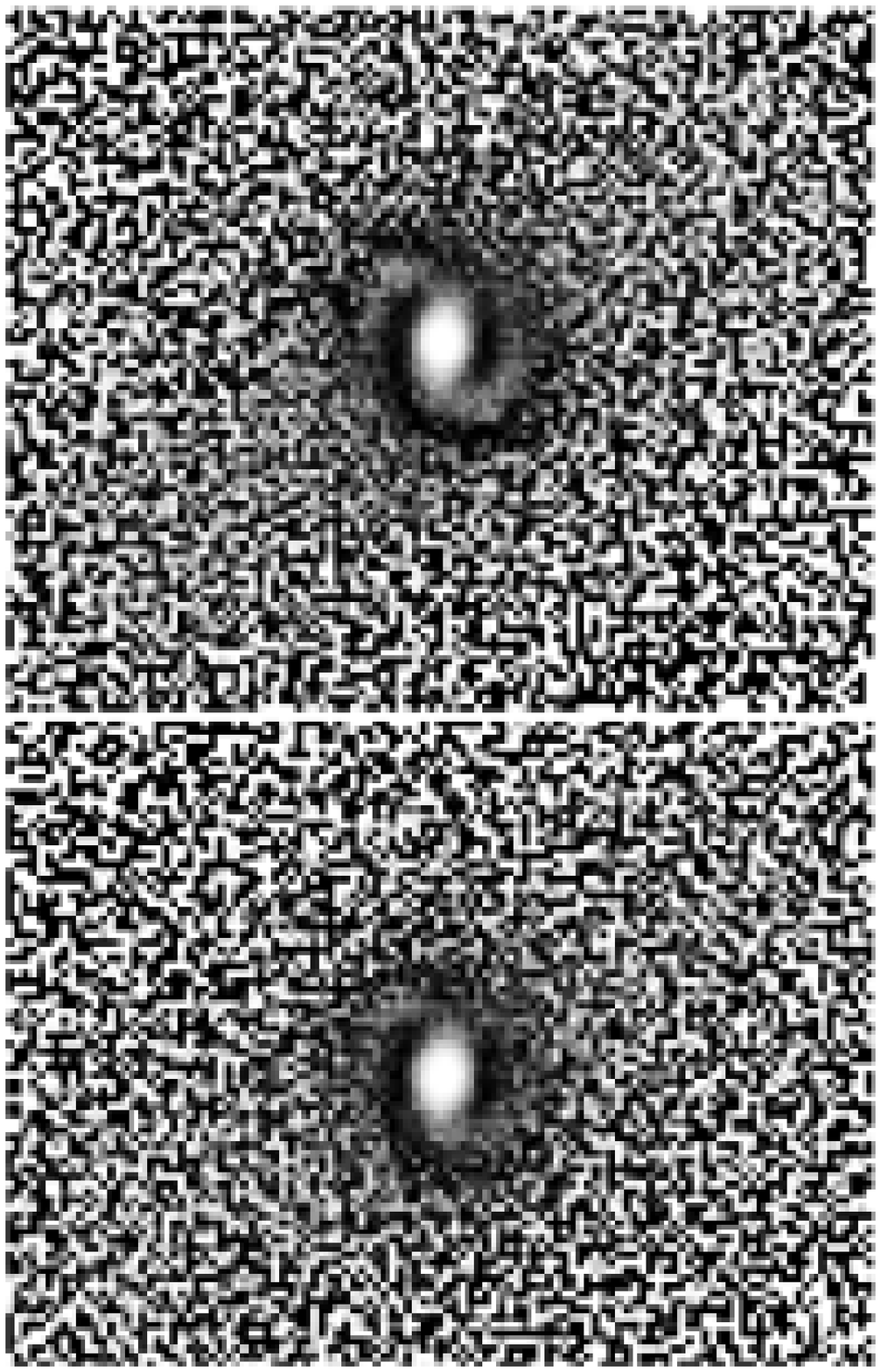}
\includegraphics[width=80mm]{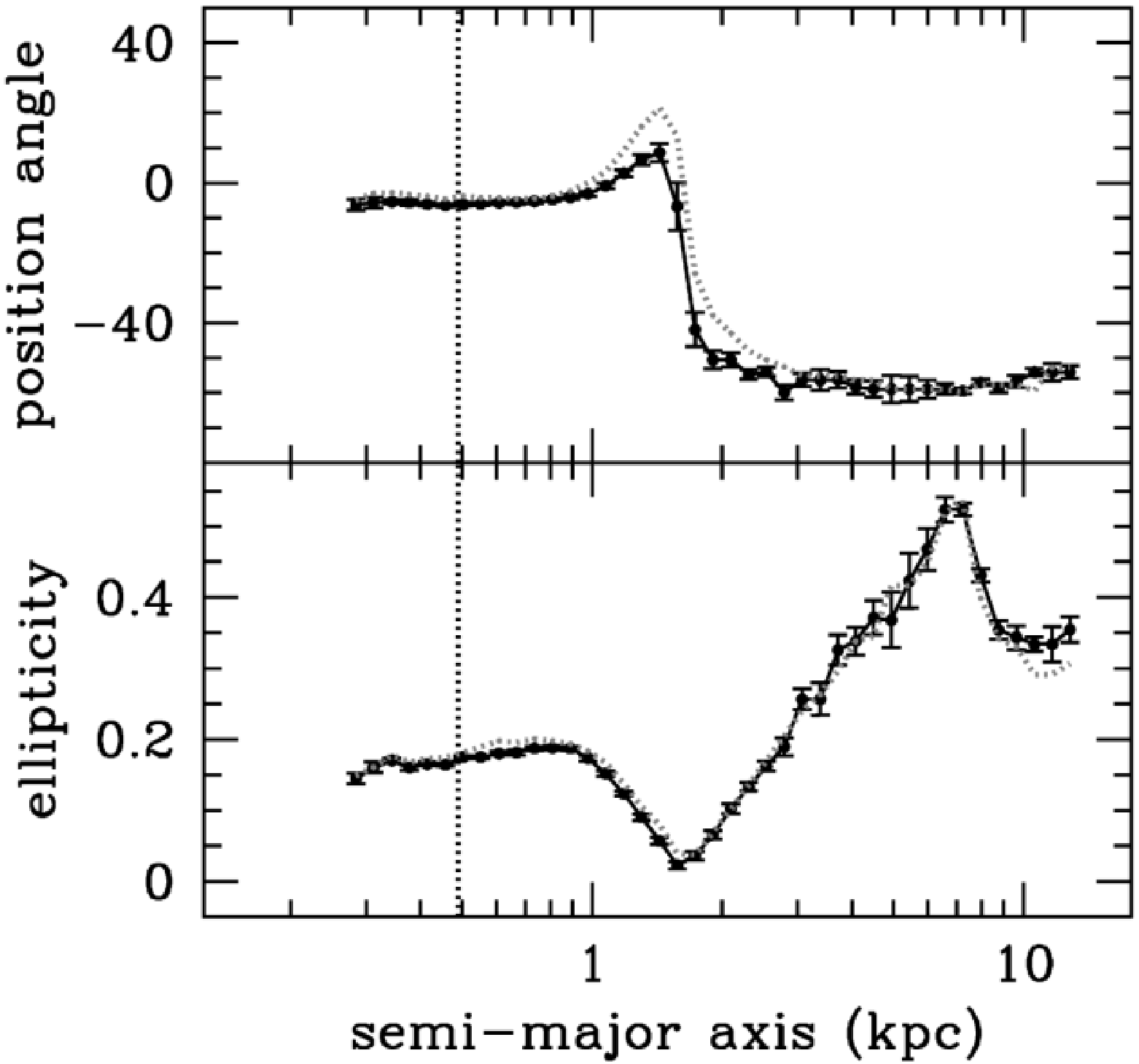}
}
\caption{{\bf Simulated observation of S2B\,1 at $z=0.3$.} Unsharp mask
  ($\sigma=1.5\,\rm{pix}$) and ellipse fits of S2B\,1 after artificially
  redshifting it to $z=0.3$. Both $V$-band (upper mask image; dotted
  grey line) and $i$-band (lower mask image; solid line) are shown. The
  dotted vertical line indicates the value of 1 PSF FWHM
  ($0\farcs11$).
 The
  corresponding Gini profile is included in Figure
  \ref{fig:mopro}. See text for details.}
\label{fig:barryat03}
\end{figure*}

Another important result for future searches of S2Bs stems from our
detection of S2B\,2: its inner bar has an apparent semi-major axis
size of only $a_{\epsilon,2}=1.5$ PSF FWHM but can still be
recognized. This is significantly smaller than the estimate of $2.5$ PSF
FWHM as a lower limit from \citet{she03}, probably also due to the
increased resolution from image drizzling. We thus suggest 1.5 PSF
FWHM as a more modest minimum size for detecting nuclear bars. When we
adopt this new limit, Fig.~\ref{fig:sizes} demonstrates that we
increase the maximum redshift for detection of the largest inner bars
($a_{\epsilon,2}\ga 1\,\rm{kpc}$) of local S2Bs by more than a factor
of two, from a redshift of $z\approx 0.25$ to $z\approx 0.5$. In
particular, S2B\,1 with $a_{\epsilon,2}=0\farcs3=2.7$ PSF FWHM should
still be detectable at $z\approx 0.3$ given the new detection limit.
To test this, we created a mock image of S2B\,1 at $z=0.3$. Taking
into account the corresponding wavelength shift would require an image
right between $V$-band and $i$-band to be 'redshifted'. We therefore
performed the artificial redshifting for both $V$- and $i$-band
images. The images were rescaled and convolved with a Moffat profile
representing the PSF of the GOODS images. Gaussian noise corresponding
to the noise of the original images was then added, and unsharp masks,
ellipse fits and Gini profiles were produced. The flux values were
rescaled correctly throughout the process, taking into account surface
brightness dimming as well as a proper normalization of the convolved
images. We did not deconvolve the original image prior to redshifting
it, which means that our final PSF is slightly too large; hence our
treatment is conservative.
The secondary bar can still be clearly identified in the unsharp masks
(Fig.~\ref{fig:barryat03}), and the primary bar would also have been
selected from its Gini profile (dotted line in top left panel of
Fig.~\ref{fig:mopro}). The ellipse fits (Fig.~\ref{fig:barryat03})
show the inner bar as a flat peak in ellipticity, with a sharp drop
outside of it. This inner part looks very similar to the ellipse fit
of S2B\,2, for the obvious reason that both inner bars in the
redshifted S2B\,1 and in S2B\,2 are close to their detection
limit. Since the inner bar of S2B\,1 is not exceptional compared to
local S2Bs (24\% of the inner bars listed in \citet{erw04} are even
larger), those could therefore be detected out to still larger
redshifts, $z\approx 0.5$ (Fig.~\ref{fig:sizes}).

Our identification of two S2Bs and five more S2B-candidate objects in
the COSMOS data proves that in present-day deep surveys, it is
worthwhile to search for distant S2Bs.  These galaxies can be analyzed
in a similar way as presented in this paper
as soon as redshifts and colour data become available.
We have shown that {\it
HST}/ACS data are capable of making considerable progress in
understanding the structure and evolution of such objects. We
therefore see the possibility of an S2B sample distributed over a
large redshift range in the near future.  Such a sample would serve as
an important observational constraint on models of S2B formation.

We described a first step in developing a model-independent method for
detecting large and strong bars which we used here.  While the method
as presented here is not optimized, it appears that with further
refinements it can become a useful method for identifying bars in
large surveys.

\section*{Acknowledgments}
We thank the referee for useful comments that helped us improve the
paper. We are grateful to Shanil Virani for providing us with an upper
limit on the X-ray flux of S2B\,1.
TL, VPD and IF enjoyed the company of the Celeste
family and the atmosphere at Barry's Pizza. VPD and IF thank the
Astronomisches Institut der Universit\"at Basel for hospitality while
this paper was in progress. TL gratefully acknowledges support by the
Swiss National Science Foundation through grant number 200020-105260.
  VPD is supported by a Brooks Prize
Fellowship at the University of Washington and receives partial
support from NSF ITR grant PHY-0205413.
This research has made use of NASA's Astrophysics Data System
Bibliographic Services.



\begin{thebibliography}{}

\bibitem[\protect\citeauthoryear{{Abraham}, {Merrifield}, {Ellis}, {Tanvir} \&
  {Brinchmann}}{{Abraham} et~al.}{1999}]{abr99}
{Abraham} R.~G.,  {Merrifield} M.~R.,  {Ellis} R.~S.,  {Tanvir} N.~R.,
  {Brinchmann} J.,  1999, \mnras, 308, 569

\bibitem[\protect\citeauthoryear{{Abraham}, {van den Bergh} \&
  {Nair}}{{Abraham} et~al.}{2003}]{abr03}
{Abraham} R.~G.,  {van den Bergh} S.,    {Nair} P.,  2003, \apj, 588, 218

\bibitem[\protect\citeauthoryear{{Afonso}, {Mobasher}, {Koekemoer}, {Norris} \&
  {Cram}}{{Afonso} et~al.}{2006}]{afo06}
{Afonso} J.,  {Mobasher} B.,  {Koekemoer} A.,  {Norris} R.~P.,    {Cram} L.,
  2006, AJ, 131, 1216

\bibitem[\protect\citeauthoryear{{Alexander}, {Bauer}, {Brandt} \& {et
  al.}}{{Alexander} et~al.}{2003}]{ale03}
{Alexander} D.~M.,  {Bauer} F.~E.,  {Brandt} W.~N.,    {et al.} 2003, \aj, 126,
  539

\bibitem[\protect\citeauthoryear{{Bershady}, {Jangren} \&
  {Conselice}}{{Bershady} et~al.}{2000}]{ber00}
{Bershady} M.~A.,  {Jangren} A.,    {Conselice} C.~J.,  2000, \aj, 119, 2645

\bibitem[\protect\citeauthoryear{{Bruzual} \& {Charlot}}{{Bruzual} \&
  {Charlot}}{2003}]{bc03}
{Bruzual} G.,  {Charlot} S.,  2003, \mnras, 344, 1000

\bibitem[\protect\citeauthoryear{{Buta} \& {Combes}}{{Buta} \&
  {Combes}}{1996}]{but96}
{Buta} R.,  {Combes} F.,  1996, Fundamentals of Cosmic Physics, 17, 95

\bibitem[\protect\citeauthoryear{{Buta} \& {Crocker}}{{Buta} \&
  {Crocker}}{1991}]{but91}
{Buta} R.,  {Crocker} D.~A.,  1991, \aj, 102, 1715

\bibitem[\protect\citeauthoryear{{Cappellari} \& {Copin}}{{Cappellari} \&
  {Copin}}{2003}]{cap03}
{Cappellari} M.,  {Copin} Y.,  2003, \mnras, 342, 345

\bibitem[\protect\citeauthoryear{{Charlot} \& {Fall}}{{Charlot} \&
  {Fall}}{2000}]{cha00}
{Charlot} S.,  {Fall} S.~M.,  2000, \apj, 539, 718

\bibitem[\protect\citeauthoryear{{Colless}, {Dalton}, {Maddox} \& {et
  al.}}{{Colless} et~al.}{2001}]{2df}
{Colless} M.,  {Dalton} G.,  {Maddox} S.,    {et al.} 2001, \mnras, 328, 1039

\bibitem[\protect\citeauthoryear{{Conselice}}{{Conselice}}{2003}]{con03}
{Conselice} C.~J.,  2003, \apjs, 147, 1

\bibitem[\protect\citeauthoryear{{Corsini}, {Debattista} \&
  {Aguerri}}{{Corsini} et~al.}{2003}]{cor03} {Corsini} E.~M.,
  {Debattista} V.~P., {Aguerri} J.~A.~L., 2003, \apjl, 599, L29

\bibitem[\protect\citeauthoryear{{Debattista}, {Mayer}, {Carollo},
  {Moore}, {Wadsley} \& {Quinn}}{{Debattista} et~al.}{2006}]{deb05b}
  {Debattista} V.~P., {Mayer} L., {Carollo} C.~M., {Moore} B.,
  {Wadsley} J., {Quinn} T., 2006, \apj, {\it accepted}
  (astro-ph/0509310)

\bibitem[\protect\citeauthoryear{{de Vaucouleurs}}{{de
Vaucouleurs}}{1975}]{dv75} {de Vaucouleurs} G., 1975, \apjs, 29, 193

\bibitem[\protect\citeauthoryear{{Elmegreen}, {Elmegreen} \&
  {Hirst}}{{Elmegreen} et~al.}{2004}]{elm04} {Elmegreen} B.~G.,
  {Elmegreen} D.~M., {Hirst} A.~C., 2004, \apj, 612, 191

\bibitem[\protect\citeauthoryear{{Erwin}}{{Erwin}}{2004}]{erw04}
{Erwin} P.,  2004, \aap, 415, 941

\bibitem[\protect\citeauthoryear{{Erwin} \& {Sparke}}{{Erwin} \&
  {Sparke}}{1999}]{erw99}
{Erwin} P.,  {Sparke} L.~S.,  1999, \apjl, 521, L37

\bibitem[\protect\citeauthoryear{{Erwin} \& {Sparke}}{{Erwin} \&
  {Sparke}}{2002}]{erw02}
{Erwin} P.,  {Sparke} L.~S.,  2002, \aj, 124, 65

\bibitem[\protect\citeauthoryear{{Erwin} \& {Sparke}}{{Erwin} \&
  {Sparke}}{2003}]{erw03}
{Erwin} P.,  {Sparke} L.~S.,  2003, \apjs, 146, 299

\bibitem[\protect\citeauthoryear{{Eskridge}, {Frogel}, {Pogge} \& {et
  al.}}{{Eskridge} et~al.}{2000}]{esk00}
{Eskridge} P.~B.,  {Frogel} J.~A.,  {Pogge} R.~W.,    {et al.} 2000, \aj, 119,
  536

\bibitem[\protect\citeauthoryear{{Ferreras}, {Lisker}, {Carollo}, {Lilly} \&
  {Mobasher}}{{Ferreras} et~al.}{2005}]{egoodss}
{Ferreras} I.,  {Lisker} T.,  {Carollo} C.~M.,  {Lilly} S.~J.,    {Mobasher}
  B.,  2005, \apj, 635, 243

\bibitem[\protect\citeauthoryear{{Giacconi}, {Zirm}, {Wang} \& {et
  al.}}{{Giacconi} et~al.}{2002}]{gia02}
{Giacconi} R.,  {Zirm} A.,  {Wang} J.,    {et al.} 2002, \apjs, 139, 369

\bibitem[\protect\citeauthoryear{{Giavalisco}, {Ferguson}, {Koekemoer} \& {et
  al.}}{{Giavalisco} et~al.}{2004}]{goods}
{Giavalisco} M.,  {Ferguson} H.~C.,  {Koekemoer} A.~M.,    {et al.} 2004,
  \apjl, 600, L93

\bibitem[\protect\citeauthoryear{{Gini}}{{Gini}}{1912}]{gini}
{Gini} C.,  1912, reprinted in Memorie di Metodologia Statistica, ed. E.
  Pizetti \& T. Salvemini (1955; Rome: Libreria Eredi Virgilio Veschi)

\bibitem[\protect\citeauthoryear{{Glassman}, {Larkin} \& {Lafreni{\`
  e}re}}{{Glassman} et~al.}{2002}]{gla02}
{Glassman} T.~M.,  {Larkin} J.~E.,    {Lafreni{\` e}re} D.,  2002, \apj, 581,
  865

\bibitem[\protect\citeauthoryear{{Hornschemeier}, {Bauer}, {Alexander} \& {et
  al.}}{{Hornschemeier} et~al.}{2003}]{hor03}
{Hornschemeier} A.~E.,  {Bauer} F.~E.,  {Alexander} D.~M.,    {et al.} 2003,
  \aj, 126, 575

\bibitem[\protect\citeauthoryear{{Jogee}, {Barazza}, {Rix} \& {et al.}}{{Jogee}
  et~al.}{2004}]{jog04}
{Jogee} S.,  {Barazza} F.~D.,  {Rix} H.-W.,    {et al.} 2004, \apjl, 615, L105

\bibitem[\protect\citeauthoryear{{Knapen}, {Shlosman} \& {Peletier}}{{Knapen}
  et~al.}{2000}]{kna00}
{Knapen} J.~H.,  {Shlosman} I.,    {Peletier} R.~F.,  2000, \apj, 529, 93

\bibitem[\protect\citeauthoryear{{Laine}, {Shlosman}, {Knapen} \&
  {Peletier}}{{Laine} et~al.}{2002}]{lai02}
{Laine} S.,  {Shlosman} I.,  {Knapen} J.~H.,    {Peletier} R.~F.,  2002, \apj,
  567, 97

\bibitem[\protect\citeauthoryear{{Le F{\` e}vre}, {Vettolani}, {Paltani} \& {et
  al.}}{{Le F{\` e}vre} et~al.}{2004}]{vvds}
{Le F{\` e}vre} O.,  {Vettolani} G.,  {Paltani} S.,    {et al.} 2004, \aap,
  428, 1043

\bibitem[\protect\citeauthoryear{{Lilly}, {Schade}, {Ellis} \& {et
  al.}}{{Lilly} et~al.}{1998}]{lil98}
{Lilly} S.,  {Schade} D.,  {Ellis} R.,    {et al.} 1998, \apj, 500, 75

\bibitem[\protect\citeauthoryear{{Lotz}, {Primack} \& {Madau}}{{Lotz}
  et~al.}{2004}]{lot04}
{Lotz} J.~M.,  {Primack} J.,    {Madau} P.,  2004, \aj, 128, 163

\bibitem[\protect\citeauthoryear{{Lynden-Bell} \& {Kalnajs}}{{Lynden-Bell} \&
  {Kalnajs}}{1972}]{lyn72}
{Lynden-Bell} D.,  {Kalnajs} A.~J.,  1972, \mnras, 157, 1

\bibitem[\protect\citeauthoryear{{Maciejewski} \& {Sparke}}{{Maciejewski} \&
  {Sparke}}{1997}]{mac97}
{Maciejewski} W.,  {Sparke} L.~S.,  1997, \apjl, 484, L117

\bibitem[\protect\citeauthoryear{{Maciejewski} \& {Sparke}}{{Maciejewski} \&
  {Sparke}}{2000}]{mac00}
{Maciejewski} W.,  {Sparke} L.~S.,  2000, \mnras, 313, 745

\bibitem[\protect\citeauthoryear{{P{\' e}rez}}{{P{\' e}rez}}{2004}]{per04}
{P{\' e}rez} I.,  2004, \aap, 427, L17

\bibitem[\protect\citeauthoryear{{Petrosian}}{{Petrosian}}{1976}]{pet76}
{Petrosian} V.,  1976, \apjl, 209, L1

\bibitem[\protect\citeauthoryear{{Rautiainen}, {Salo} \&
  {Laurikainen}}{{Rautiainen} et~al.}{2002}]{rau02}
{Rautiainen} P.,  {Salo} H.,    {Laurikainen} E.,  2002, \mnras, 337, 1233

\bibitem[\protect\citeauthoryear{{Reshetnikov}, {Battaner}, {Combes} \& {Jim{\'
  e}nez-Vicente}}{{Reshetnikov} et~al.}{2002}]{res02}
{Reshetnikov} V.,  {Battaner} E.,  {Combes} F.,    {Jim{\' e}nez-Vicente} J.,
  2002, \aap, 382, 513

\bibitem[\protect\citeauthoryear{{Rix}, {Barden}, {Beckwith} \& {et al.}}{{Rix}
  et~al.}{2004}]{rix04}
{Rix} H.-W.,  {Barden} M.,  {Beckwith} S.~V.~W.,    {et al.} 2004, \apjs, 152,
  163

\bibitem[\protect\citeauthoryear{{Sellwood} \& {Wilkinson}}{{Sellwood}
  \& {Wilkinson}}{1993}]{sw93review} {Sellwood} J.~A., {Wilkinson} A.,
  1993, Rep. Prog. Phys, 56, 173

\bibitem[\protect\citeauthoryear{{Sheth}, {Regan}, {Scoville} \&
  {Strubbe}}{{Sheth} et~al.}{2003}]{she03}
{Sheth} K.,  {Regan} M.~W.,  {Scoville} N.~Z.,    {Strubbe} L.~E.,  2003,
  \apjl, 592, L13

\bibitem[\protect\citeauthoryear{{Shlosman}, {Begelman} \& {Frank}}{{Shlosman}
  et~al.}{1990}]{shl90}
{Shlosman} I.,  {Begelman} M.~C.,    {Frank} J.,  1990, \nat, 345, 679

\bibitem[\protect\citeauthoryear{{Shlosman}, {Frank} \& {Begelman}}{{Shlosman}
  et~al.}{1989}]{shl89}
{Shlosman} I.,  {Frank} J.,    {Begelman} M.~C.,  1989, \nat, 338, 45

\bibitem[\protect\citeauthoryear{{Simard}, {Koo}, {Faber} \& {et al.}}{{Simard}
  et~al.}{1999}]{sim99}
{Simard} L.,  {Koo} D.~C.,  {Faber} S.~M.,    {et al.} 1999, \apj, 519, 563

\bibitem[\protect\citeauthoryear{{Trujillo} \& {Pohlen}}{{Trujillo} \&
  {Pohlen}}{2005}]{tru05}
{Trujillo} I.,  {Pohlen} M.,  2005, \apjl, 630, L17

\bibitem[\protect\citeauthoryear{{Virani}, {Treister}, {Urry} \&
  {Gawiser}}{{Virani} et~al.}{2006}]{vir06}
{Virani} S.,  {Treister} E.,  {Urry} C.~M.,    {Gawiser} E.,  2006, AJ, {\it accepted} (astro-ph/0506551)

\bibitem[\protect\citeauthoryear{{Wolf}, {Meisenheimer}, {Kleinheinrich},
  {Borch}, {Dye}, {Gray}, {Wisotzki}, {Bell}, {Rix}, {Cimatti}, {Hasinger} \&
  {Szokoly}}{{Wolf} et~al.}{2004}]{combo17}
{Wolf} C.,  {Meisenheimer} K.,  {Kleinheinrich} M.,  {Borch} A.,  {Dye} S.,
  {Gray} M.,  {Wisotzki} L.,  {Bell} E.~F.,  {Rix} H.-W.,  {Cimatti} A.,
  {Hasinger} G.,    {Szokoly} G.,  2004, \aap, 421, 913

\end{thebibliography}


\begin{table*}
\begin{minipage}{126mm}
\caption{{\bf Photometric properties of S2B\,1 and
    S2B\,2.} All magnitudes and colours are given in the AB system.
 Magnitude errors are estimated to be $0\fm1$.
 Formal errors from the SNR are much smaller.
 Similarly, formal errors on the colour values are negligible. Instead,
    we refer the reader to Figs.\ \ref{fig:BC} and \ref{fig:MC}
    concerning the colour \emph{variation} within the galaxies.
See text for further details.
}
\label{tab:phot}
\begin{tabular}{lllllll}
Object                    & ID     & m$_{\rm i}$    & M$_{\rm V}$    & $B-V$          & $V-i$          & $i-z$             \\
                          &        & mag$_{\rm AB}$ & mag$_{\rm AB}$ & mag$_{\rm AB}$ & mag$_{\rm AB}$ & mag$_{\rm AB}$    \\
\hline
GOODS J033230.93-273923.7 & S2B\,1 & $17.9$         & $-20.8$        & ---            & $0.57$         & $0.27$  \\
GOODS J033233.46-274312.8 & S2B\,2 & $17.1$         & $-20.6$        & $1.31$         & $0.57$         & $0.28$  \\ 
\hline
\end{tabular}
\end{minipage}
\end{table*}

\begin{table*}
\begin{minipage}{126mm}
\caption{{\bf S2Bs and S2B candidates in COSMOS.}
 Magnitude errors are estimated to be $0\fm1$.
}
\label{tab:cosmos}
\begin{tabular}{lll}
Object              & m$_{\rm i}$ & Classification\\
                    & mag$_{\rm AB}$              &               \\
\hline
COSMOS J095922.76+024245.3 & 18.4             &  S2B \\ 
COSMOS J095936.98+015107.6 & 17.9             &  S2B \\ 
COSMOS J095931.25+015916.5 & 18.1$^a$         &  possibly S2B or SB with nuclear disk \\ 
COSMOS J095947.65+021549.0 & 18.4             &  possibly S2B or SB with nuclear disk \\ 
COSMOS J100045.42+021932.4 & 19.4             &  possibly S2B or SB with nuclear disk \\ 
COSMOS J100048.23+015856.2 & 20.0             &  possibly S2B or SB with nuclear disk \\ 
COSMOS J095952.25+020543.1 & 20.1             &  possibly S2B \\ 
\hline
\end{tabular}
 \begin{list}{}{}
 \item[$^{\rm{a}}$] Partially contaminated by a
 bright star's bleeding trails.
 \end{list}
\end{minipage}
\end{table*}

\bsp

\label{lastpage}

\end{document}